\documentclass[prd,tightenlines,twocolumn]{revtex4}
\usepackage{amsmath}
\usepackage{amssymb}
\usepackage{graphicx}
\usepackage{bm}
\usepackage{enumerate}
\usepackage{color}
\newcommand{\be}{\begin{equation}}
\newcommand{\ee}{\end{equation}}
\newcommand{\bea}{\begin{eqnarray}}
\newcommand{\eea}{\end{eqnarray}}
\newcommand{\ba}{\begin{eqnarray}}
\newcommand{\ea}{\end{eqnarray}}

\begin{document}

\title{ High Multiplicity pp and pA Collisions: \\ The Hydrodynamics at its Edge }
\author{ 
 Edward Shuryak and Ismail Zahed}

\affiliation{Department of Physics and Astronomy, \\ Stony Brook University,\\
Stony Brook, NY 11794, USA}

\date{\today}

\begin{abstract} 
With growing multiplicity, the pp and pA collisions enter the domain where the macroscopic description (thermodynamics and hydrodynamics) becomes applicable. We discuss this situation, first with simplified thought experiments, then
with some idealized representative cases, and finally address the real data.
For clarity, we don't do it numerically but analytically, using the Gubser solution. We found that the radial flow is expected to increase from central AA to central pA, while
the elliptic flow decreases, with higher harmonics being comparable. 
In the second part of the paper we approach the problem from the opposite side, using a string-based Pomeron model.
We extensively study the magnitude and distribution of the viscous corrections, in Navier-Stokes and Israel-Stuart
approximations, ending with higher gradient re-summation proposed by Lublinsky and Shuryak. We found
those corrections growing, from AA to pA to pp, but remaining at the manageable size even in the last case. 
 \end{abstract}

\maketitle
\section{Introduction}
High energy heavy ion collisions are theoretically treated very differently from pp and pA ones. While the former are very well described using macroscopic theories -- thermodynamics and relativistic hydrodynamics -- the latter are subject to what we would like to call the ``pomeron physics", described with a help of microscopic dynamics in terms of (ladders of) perturbative gluons, classical random gauge fields, or strings.
The temperature and entropy play a central role in the former case, and are not even mentioned or defined in the latter case. 

The subject of this paper is the situation when these two distinct worlds (perhaps) meet. In short, the main statement of this paper is that
specially triggered fluctuations of the pp and pA collisions of particular magnitude should be able to reach conditions in which
the macroscopic  description can be nearly as good as for AA collisions. While triggered by experimental hints at LHC to be discussed below,
this phenomenon has not yet been a subject of a systematic study experimentally or theoretically, and is of course far from
being understood.  So on onset let us enumerate few key issues to be addressed. 

\begin{itemize}

\item  How do the thermodynamical and hydrodynamical (viscosities, relaxation time etc) quantities scale with the change in the system size $R$ and the multiplicity $N$? What are the criteria for macroscopic  
(hydrodynamical) behavior ?

\item What are the consequences of the fact that the sQGP phase of matter is approximately scale invariant ?

\item Do high multiplicity pp and pA  collisions in which the (double) ``ridge" has been recently observed
at LHC \cite{CMS:2012qk,Abelev:2012cya,ATLAS} fit into the hydrodynamical systematics tested so far
for AA collisions?

\item What is the expected magnitude of the radial flow in  pp and pA  collisions, and how is it related to that in AA?
What are the freezeout conditions in these new explosive systems?

\item How do amplitudes of  the second and higher angular harmonics $v_n$ scale with $n$,$R$ and $\eta/s$?
In which $p_t$ region do we expect hydrodynamics to work, and for with $v_n$?

 \end{itemize}

   
    The major objective of the heavy ion collision program is to create and study properties
    of a new form of matter, the Quark-Gluon Plasma. Among many proposed signatures proposed
   in \cite{Shuryak:1978ij}, the central role is played by production of macroscopic fireball of such matter, with the
subsequent   collective explosion described by the relativistic hydrodynamics.
  Its observable effects are include radial and elliptic flow, supplemented by higher moments $v_m,m>2$ .
 At RHIC and LHC the  AA collisions has been studied in detail by now, with multiple measured dependences, with
   excellent agreement
 with hydrodynamics in a wide domain, for $n<7$ and 
 in the range of $p_t< 3 \,GeV$.  
 
    Let us start with a very generic discussion of applicability of hydrodynamics.  The basic condition is that the system's size $R$ should be much larger than microscopic
scales such as e.g. the correlation lengths or the inverse temperature $T^{-1}$. The corresponding ratio is one small parameter 
\ba  \label{eqn_par1}
 {1 \over T R} \approx {\cal O}(1/10) \ll 1
 \ea
 where the value corresponds to well studied central AA collisions.
Another important small parameter which we seem to have for strongly coupled Quark-Gluon Plasma (sQGP) is the {\em viscosity-to-entropy-density ratio} 
\be {\eta \over s} =0.1..0.2 \ll 1    \label{eqn_par2}\ee
 This tells us that viscous scale -- the mean free path in kinetic terms -- is additionally suppressed compared to the
 micro scale $1/T$ by strong interaction in the system. The {\em product} of both parameters appearing
in  expressions (to be specified below)  
  suggests that one can hope to apply
 hydrodynamics with about percent accuracy. 
 
\begin{figure}[t]
\begin{center}
\includegraphics[width=7cm]{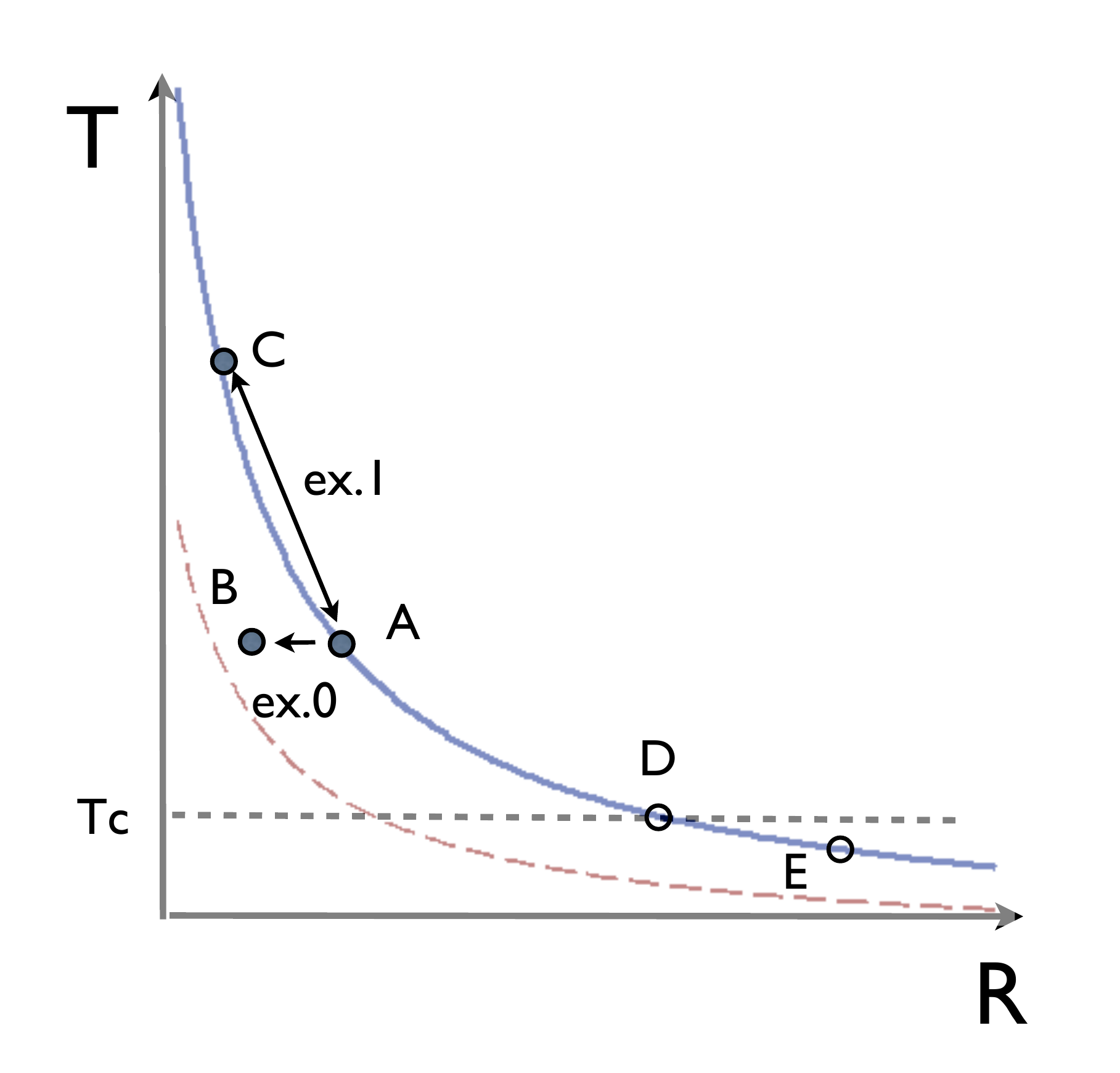}
\caption{(color online)  Temperature $T$ versus the fireball size $R$ plane. Solid blue line is the adiabate $S=const$,
approximately $TR=const$ for sQGP. Example 0 in the text corresponds to reducing $R$, moving left $A \rightarrow B$.
Example 1 is moving up the adiabate $A \rightarrow C$.  Example 2 corresponds to adiabatic expansion, such as
$A \rightarrow E$,$C \rightarrow E$. If in reality $C$ corresponds to $pA$, the freezeout occurs at the earlier point $D$.      }
\label{fig_TR}
\end{center}
\end{figure}

     The reason  why the  fireballs produced in AuAu collisions at RHIC and PbPb at LHC behaves macroscopically is
related to the   {\em large size} of the colliding nuclei used. Yet
    smaller size  systems
   occurring in pp or pA should also be able to do so, provided certain conditions are met. 
      Let us thus start to define such a comparison, starting with our {\em thought experiment 0}, in which two systems 
      (see a sketch in Fig.\ref{fig_TR}) $A$ and $B$
      have the same
local quantities -- temperatures, viscosities and the like -- but different sizes $R_A  > R_B$.  
(For example, think of AuAu and CuCu collisions at the same collision energy,  as in experiments done at RHIC.)  
    The equations of {\em ideal} hydrodynamics 
\be \partial_\mu T^{\mu\nu}=0 \ee
include derivatives   linearly
and therefore simultaneous rescaling of the 
size and the  time $x^\mu \rightarrow \lambda x^\mu$ does not change them. So, $ideal$ hydrodynamics
will produce the same solution for fireball of any size, provided other parameters are unchanged.
Yet the viscous terms have more gradients, and thus there is no such symmetry. Going from
   a large AA fireball to smaller $pA .. pp$ systems would increases the role of visous terms (scaled as powers of $1/R$) , eventually invalidating hydrodynamics.  (The boundary of which is shown in Fig.\ref{fig_TR} by red long-dashed line.)
     
 However
if local quantities such as $T$ are changed as well, as is indeed the case in experimental conditions we will discuss, the conclusion may change.      
     Consider instead  the {\em thought experiment 1}, in which we compare two systems 
     on the same adiabate $A$ and $C$.  For conformally invariant  sQGP  -- such as exists in the
   $\cal{N}$=4 supersymmetric theory without  running coupling -- $S\sim (TR)^3=const$ and the points  $A,C$ are
  related by the scale transformation  \be R_A/R_C=\xi,T_A/T_C=  \xi^{-1} \ee
    If the scale transformation is a symmetry,
   all densities --  e.g. the energy densities -- scale with
   the naive dimensional powers of the temperature $ \epsilon/T^4\sim const$, viscosities do the same. 
    Thus {\em the absolute scale  plays no role}.      A small (but hotter) plasma
   ball $C$ will behave exactly in the same way as the large (but cooler) $A$, provided all dimensionless quantities like $TR$ 
   or total entropy/multiplicity are held constant.     

Let us now proceed to the {\em thought experiment 2}, which is the same as above but in QCD, with a running coupling.
In the sQGP regime it leads to (very small, as lattice tells us ) running of $s/T^3$, of (unknown) running of $\eta/T^3$ etc. 
The most dramatic effect is not the running coupling {\em per se}, but the lack of supersymmetry,
which induces  chiral/deconfinement phase transition out of the sQGP
phase at $T=T_c$.  The end of the sQGP explosion $D$ thus has an {\em absolute scale}, not subject to scale transformation!

So let us consider two systems $A$,$C$ of the same total entropy/multiplicity, initiated in sQGP with conditions related by scale transformation
and left  them explode.
The sQGP evolution would be related by nearly the same set of intermediate states (modulo running coupling)  till $T\approx T_c$,
after which they go into the ``mixed" and hadronic stages, which are $not$ even close to be scale invariant!  Thus the 
result of the explosions are not the same. In fact the smaller/hotter system will have an advantage over the larger/cooler one,
since it has $larger$  ratio between the initial and final scales $T_i/T_f$. 

(In the language of holographic models the scale is interpreted as the 5-th coordinate $x^5$, and evolution is depicted as gravitational falling of particles,strings, fireballs etc toward the AdS center. The ratio of the scales is the distance travelled in the
5-th coordinate: thus in this language two systems fall similarly in the same gravity, but smaller system starts ``higher" and thus got
larger velocity at the same level given by $T_c$.) 

The hydro expansion does not need to stop at the phase boundary $D$. In fact large systems, as obtained in central AA
collisions are known to freezeout at $T_f<T_c$, down to 100 \, MeV range (and indicated in the sketch by the point $E$. 
However small systems, obtained in peripheral AA or central pA seem to freezeout at $D$, as we will show at the end of the paper.

Short summary of these thought experiments: not only one expects hydro in  the smaller/hotter system to be there, it should be similar
to the one in larger/cooler system, due to approximate scale invariance of sQGP. Furthermore,  in fact smaller systems are expected to produce $stronger$ hydro flow, as they evolve ``longer" (not in absolute but in dimensionalless time).

     If one wants to make comparison along such lines, the question is how one can increase the temperature of the system
     in practice. One obvious way to do so is  to increase the collision energy: taking a pair of lighter nuclei $A'A'$ at LHC one can
compare it to collision of heavier nuclei $AA$ at RHIC tuning the energy so that the multiplicity and centrality of the collisions be the same, reproducing our  thought experiment 2. Yet    
     energy dependence of multiplicity is very slow, RHIC and LHC have different detectors etc: so it is not very practical. 
           Another option is to rely on 
     rare fluctuations, selecting events with a larger entropy/multiplicity. This is very expensive\footnote{ While the cost of  an average ``min.bias" $pp$ collision is (number of events)/(cost of LHC)$\sim 10^{10}/10^{10} \$= 1\$$,   in the selected sample the cost is then about $10^6 \$ $/event.},  but
this is what is done in practice. 
     
     Let us now briefly outline the history of the subject of collective flow effects in pp collisions.
The radial flow effects in 
were searched for  in the minimum-bias pp collisions at CERN ISR more than 30 years ago by one of us \cite{Shuryak:1979ds}, with
negative results. Indications for some  radial flow
have been found in specially triggered $\bar{\rm p}$p collisions by the FERMILAB MINIMAX experiment \cite{Brooks:1999xy},
but the data remained inconclusive and, more importantly, the magnitude of the flow was small,  below 
of what the full-fledged hydro would give.  (We are not aware of any actual comparison with these data.)

    With the advent of the LHC era of extremely high luminocities and short-time detector capabilities,  a hunt for
   strong fluctuations in the parton multiplicity became possible. 
   Already during the very first run of LHC in 2010, the CMS collaboration was able \cite{cms_ridge} to
   collect sufficient sample of high multiplicity $pp$ collisions occurring with the probability $\sim 10^{-6}$. CMS found the ``ridge"
   correlation in the highest multiplicity bins, 
an angular correlation in the azimuthal angle between two particles
  at $\Delta\phi<1$ which extends to large rapidity range $|\Delta y| \geq 4$. 
 More recently the same phenomenon was seen in pPb
collisions as well, now by the CMS \cite{CMS:2012qk}, ALICE \cite{Abelev:2012cya}
 and ATLAS \cite{ATLAS} collaborations, as well as by PHENIX \cite{Adare:2013piz} in dAu collisions at RHIC.
Larger number of ``participant nucleons" and higher average multiplicity
 substantially weaken the cost of the trigger: the ``ridge" is seen at the trigger level of few percents higher multiplicity events.
It is shown in those works that in pp and pA collisions, the same threshold in terms of multiplicity is needed to start showing the ``ridge".
 
 Angular correlations naturally appear in a hydrodynamical explosion of a non-azimuthally symmetric 
 objects. The spatial shape is then translated to momentum space and is observed. For example,
  in the comments on the CMS discovery written by one of us 
  \cite{Shuryak:2010wp} it was illustrated by a string placed outside of an (axially symmetric) stick of explosive.
 While the basic wind   blowing is isotropic in $\phi$, an extra string may move in a preferred direction.
  In central AA collisions it is similar to that. A symmetric explosion has perturbations
  in the form of localized ``hot spots".  But in general, any sufficiently deformed
initial collisions for the fireball would be sufficient  to create ridge-like correlations.

Furthermore, the subtraction of the so called ``back-to back recoil" (a peak at $\phi\sim \pi$) (evaluated from
 some  perturbative  (e.g. HIJING) or color glass models \cite{KEVIN} or seen experimentally in  smaller multiplicity
bins) reveals that a ridge is ``doubledÓ on the away side. 
The remaining correlation function is  found to be \cite{Abelev:2012cya,ATLAS} nearly symmetric with  $x\rightarrow -x, \phi\rightarrow \pi-\phi$. Furthermore,  the $second$ angular harmonics  completely dominate the correlator  -- unlike the central
 AA, in which the strongest harmonics is the third.
The first attempts to describe this phenomenon hydrodynamically are qualitatively consistent with these data. For 
the pA case, it is Ref. \cite{Bozek:2011if}, which starts from Glauber-inspired initial conditions
similarly to what is done in the AA case.

 A  
nucleon propagating through the diameter of the Pb nucleus ``wounds"  up to 20 nucleons. Similar number of ``wounded nucleons" 
and multiplicity can be found for very peripheral PbPb collisions. Since these two systems have different transverse
area, they approximately correspond to our ``thought experiment 2" (modulo different shape, which can be accounted for, see below). 

The objective of this paper is to extend hydrodynamical studies, using instead of a complicated ``realistic models" with huge number of details and
heavy numerics (the ``event-by-event" hydrodynamics) an $analytic$ approach. 
 As we will see, this will allow us to focus on generic dependences of the predictions on the parameters of the problem.

The structure of the paper is as follows. In the next section we discuss the radial flow using Gubser's solution.
After putting AA,pA,pp representative cases into common dimensionless units, we see that they are in fact not so
far from thought experiments just discussed. We will then study viscous effects, from the Navier-Stokes term, 
to Israel-Stuart equations and Lublinsky-Shuryak higher gradient re-summation in section \ref{sec_higher_grad}. We found an artifact of Gubser
solution --large corrections on the space-like part of the freeze out surface, but other than that all viscous effects
seem to be reasonabley under control, in all cases considered.
We then turn to the harmonics of the flow $v_m$ in the next section, with m=2,3 and higher. We start with ``acoustic damping"
formula, outlying dependence on the parameters, and then proceed to solving the equations for Gubser flow perturbations
in AA,pA and pp cases. The last section is devoted to comparison to the experimental data. Only very recently
spectra of the identified secondaries for high-multiplicity pA had allowed to confirm our main point: the increase of the radial flow,
and even determine more quantitatively the freeze out conditions.

%

\section{Hydrodynamics of the  radial flow}

\subsection{Ideal  hydrodynamics and the Gubser's flow} 

Since we are  interested in comparison of different systems,  it is important not to have too many details which
can be different and induced some variations in both. In particularly, one should keep the matter distribution of the same shape.
It is sufficient for this purpose to use a relatively simple analytic solution found by Gubser \cite{Gubser:2010ze}, see also
\cite{Gubser:2010ui}.  This
 solution has two symmetries: the boost-invariance as well as the
axial symmetry in the transverse plane. It is obtained  via  special conformal transformation, and
therefore, the matter is required to be conformal, with the EOS 
\be \epsilon=3p =T^4 f_* \ee
where the parameter  $f_*=11$ is fitted to reproduce the lattice data on QGP thermodynamics (not too close to $T_c$).

The coordinate sets used are either the usual proper time -spatial rapidity - transverse radius - azimuthal angle
 $(\bar{\tau},\eta,\bar{r},\phi)$ set with the metric
\begin{eqnarray}
ds^2 & = & -d\bar{\tau}^2 + \bar{\tau}^2 d\eta^2 + d\bar{r}^2 +\bar{r}^2d\phi^2,
\end{eqnarray}
or the comoving coordinates we will introduce a bit later.

The shape of the solution is fixed, and the absolute scale is introduced by a single  parameter $q$ with dimension of the inverse length.
We call the
 dimensionful  variables $\bar{\tau},\bar{r}$ with the bar, which  disappears as we proceed 
to dimensionless variables
\be  t=q \bar{\tau}, \,\,\,r=q \bar{r} \ee
In such variable there is one single solution
 of ideal relativistic hydrodynamics, which for the transverse velocity and the energy density reads
\begin{eqnarray}
v_\perp(t,r) & = & \frac{2 t r }{1+t^2 + r^2} \label{eqn_ideal}
\end{eqnarray}
\begin{eqnarray}
{\epsilon \over q^4} & = & \frac{\hat{\epsilon}_0 2
^{8/3}}{t^{4/3}\left[1+2(t^2 +
r^2)+(t^2-r^2)^2\right]^{4/3}} \nonumber
\end{eqnarray}
The specificity of the  system considered is reduced to a single dimensionalless
parameter \be\hat{\epsilon}_0 = \ee 
related to macro-to-micro ratio (\ref{eqn_par1}) or multiplicity, 
plus of course different  freezeouts to which we turn shortly.

Let us crudely map the AA, pA and pp collisions to these coordinates, guessing  the scale factors in fm to be
\be q_{AA}^{-1}= 4.3 , \,\, q_{pA}^{-1}=1, \,\,\, q_{pp}^{-1}=0.5  \ee
The energy density parameter can be related to the entropy-per-rapidity density of the solution
\be \hat{\epsilon}_0=f_*^{-1/3} \left({3 \over 16 \pi} {dS \over d\eta}\right)^{4/3} \ee
which in turn is mapped to multiplicity density per unit rapidity\be {dS \over d\eta}\approx 7.5 \,\,{dN_{ch} \over d\eta} \ee
defined at freezeout.  
 We use for central  LHC AA=PbPb collisions  \be dN^{AA}_{ch}/d\eta=  1450 \ee 
The pp and pA data are split into  several multiplicity bins:  for definiteness, we will refer to  
one of them in the CMS set, with the (corrected average) multiplicity $N_{ch}=114$ inside $|\eta|<2.4$ and $p_t>0.4\, GeV$ acceptance.
We thus take 
 \be dN^{pA}_{ch}/d\eta=  dN^{pp}_{ch}/d\eta=1.6\, \,{114 \over 2*2.4} \ee
 where the factor 1.6 approximately corrects for the unobserved $p_t<0.4\, GeV$ region.
 Similarly the energy parameters are  fixed for each multiplicity bin. 
 
 (For clarity:  our thought experiments 1 and 2 of the Introduction
 assumed the $same$ values of $\hat\epsilon_0$ for points $A$ and $C$, thus the same solution. Now we compare
 central AA and some representative bins of pA and pp, which have parameters and 
 correspond  to  $different$ adiabatic curves.  )

\begin{figure}[t]
\begin{center}
\includegraphics[width=7cm]{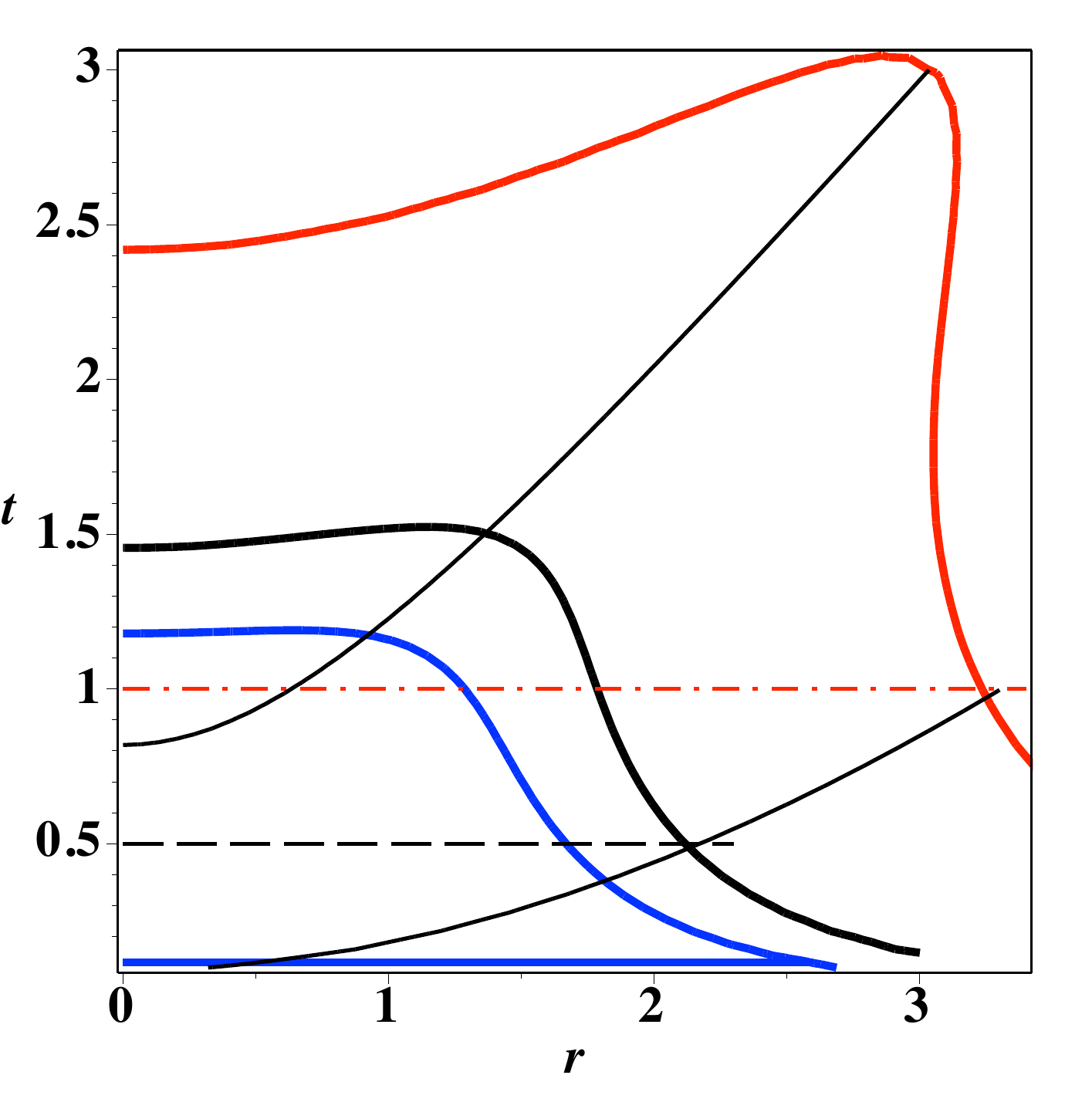}
\caption{(color online) The three horizontal lines correspond to the initial time: from bottom up AA (blue solid), pA (dash black)
 and pp (red dash-dot). The corresponding three curves with the same color 
 are the lines at which the temperature reaches the same freeze-out value, set to be
 $T_f= 150\, MeV$.  The two thin solid lines correspond to the values of the variable $\rho=-2.2$ (lower) and $-0.2$ (upper). Those
 values are used as initial and final values  in the evolution of higher harmonics.
   }
\label{fig_tr_ideal}
\end{center}
\end{figure}

    The expression for transverse flow (\ref{eqn_ideal}) does not depend on $\hat\epsilon_0$ though, and all one needs to do to calculate the radial flow is to
  define the freezeout surfaces.
 Such a map is shown on the $t,r$ plot in Fig.\ref{fig_tr_ideal}, in which we, for now, selected the same ``average" freezeout
 temperature $T_f=150 \, MeV$ (to be modified later).
 Hydrodynamics is valid between the (horizontal) initial time lines and the contours of fixed freeze out 
 temperature $T_f$, shown by thicker solid line, at which the particle decouple and fly to the detector. 
  The spectra should be calculated by the standard Cooper-Fry formula
 \be  {dN \over d\eta dp_\perp^2}\sim \int p^\mu d\Sigma_\mu\,\, {\rm exp}\left(- {p^\mu u_\mu \over T_f}\right)
 \ee
 in which $\Sigma_\mu$ is the freeze out surface, on which the collective velocity $u_\mu(t,r)$ should be taken, for details see \cite{Staig:2011wj} .
( We ignore changes in the
 equation of state at $T>T_c$.)
 
  Note first, that while the absolute sizes and multiplicities in central AA  are  quite different from
 pA and pp bins discussed, in the dimensionless variables those are not so far away. Notably
the pA freezeout appears  ``later" than for AA, and pp later still. (Of course, the order is opposite  in the absolute fm units.) 
Thus illustrates the case we made with the thought experiment 2: smaller systems gets more and more ``explosive", because in the right units $CD$ path is
longer than $AD$.

 The transverse collective velocity on the freeze out curves is read off (\ref{eqn_ideal}). We would not give here a plot but just mention that transverse rapidity rise about linearly from the fireball center to the
 the maximal values reached at the ``corner" of the freeze-out curves. For three cases considered those are
 \be v_\perp^{max} [AA, pA, pp]=[0.69, 0.83, 0.95]  \label{eqn_radial} \ee
   These values are of course for qualitative purposes only, to demonstrate the point in the most simple way. We will discuss recent CMS data and 
   realistic freezeout surfaces corresponding to them at the end of the paper.
 
\subsection{The   Navier-Stokes   corrections}
We continue to discuss the radial flow adding the first viscosity effect.
The equation for the reduced temperature $\hat{T}=\hat{\epsilon}^{1/4}$ using the combination of variables
\be g= { 1-t^2+r^2 \over 2 t}\ee
becomes an ordinary differential equation
\be 3(1+g^2)^{3/2} {d\hat{T} \over dg}+2 g \sqrt{1+g^2} \hat{T}+g^2 H_0=0 \ee
This equation is easily solvable analytically in terns of certain hypergeometric functions or numerically.
Note that the last term contains viscous parameter
\be H_0={\eta \over \epsilon^{3/4}}={\eta\over s} {4\over 3} f_*^{1/4} \ee 
For $\eta/s=0.134$ one finds $H_0=0.33$ we will use as representative number.

\begin{figure}[t]
\begin{center}
\includegraphics[width=4.5cm]{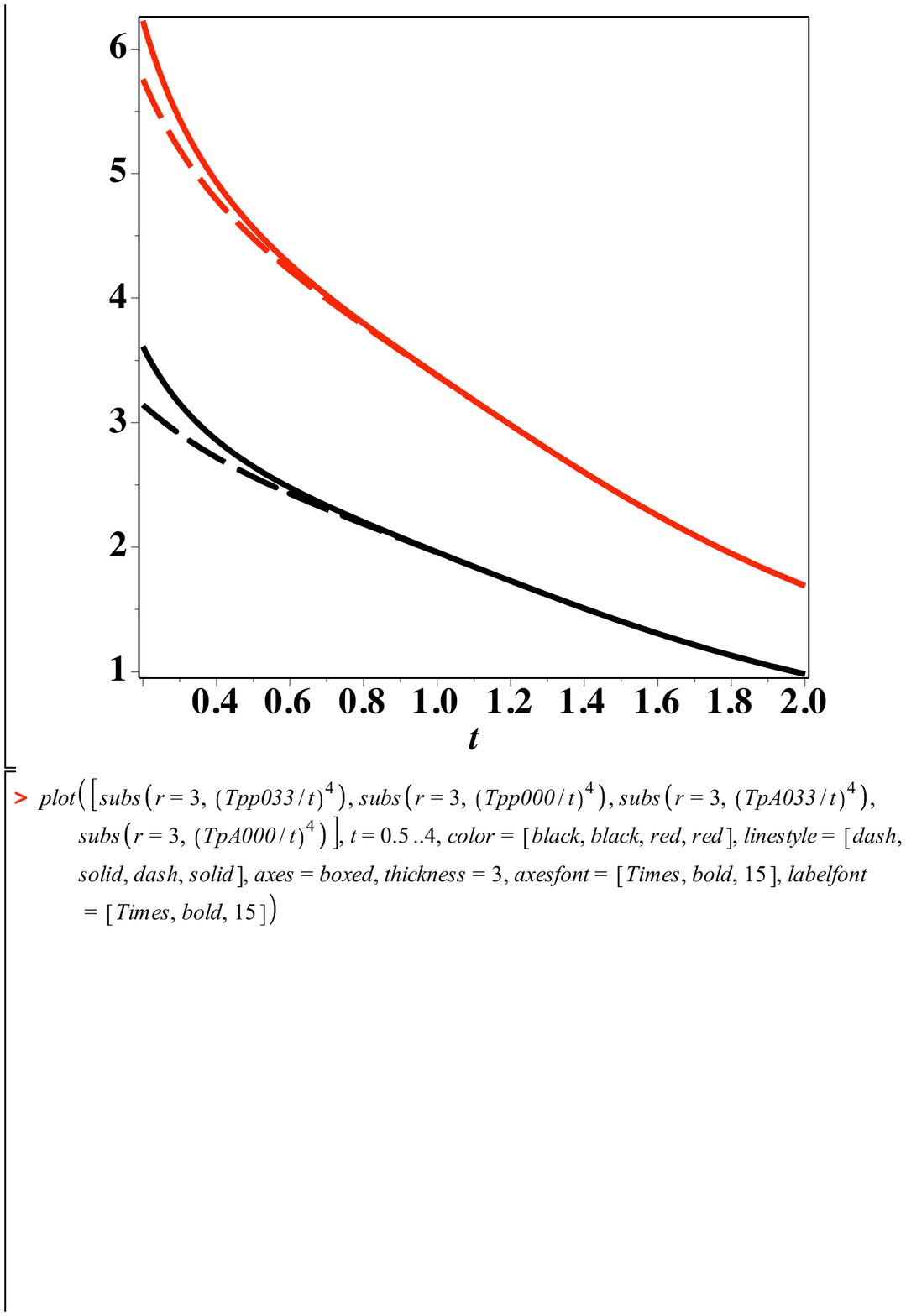}
\includegraphics[width=4.5cm]{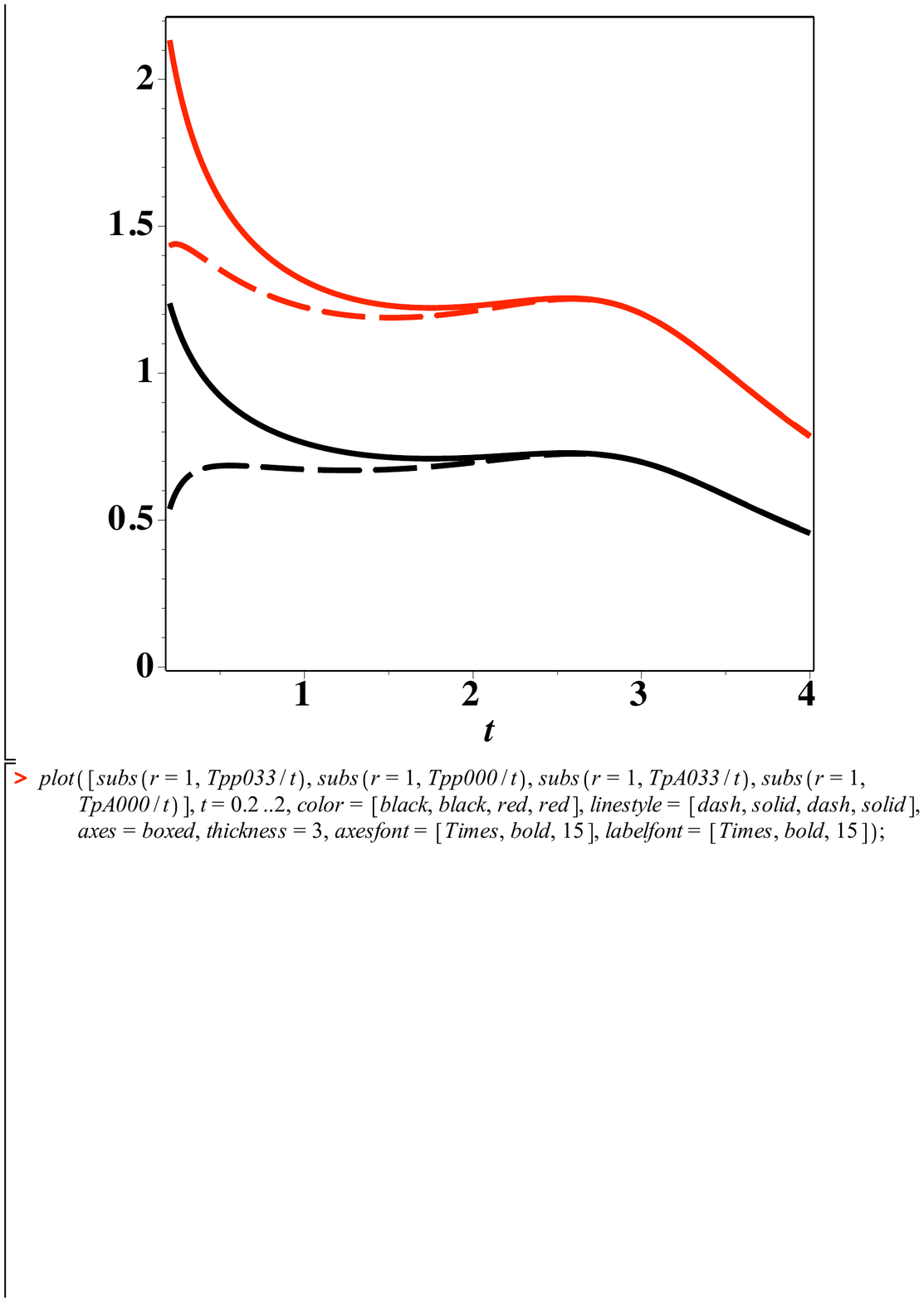}
\caption{(color online) The temperature versus dimensionless time $t$, for ideal hydrodynamics
(solid) and viscous hydrodynamics with $\eta/s=0.132$ (dashed) lines. The 
upper pair of (red) curves are for pp, the lower (black) ones for pA collisions.
The upper plot is for $r=1$, the lower plot for $r=3$.
   }
\label{fig_visc_rad}
\end{center}
\end{figure}

  The question is how important is the viscous term.
While $H_0$ is just a constant, its role  depends on the magnitude of the initial temperature $\hat{T}_0$ or total entropy. 
For AA collisions we find that its role is truly negligible, as the curves hardly are separated by the line width.  (This is, of course, well known from all studies in the literature.)
For the pA and pp cases as modeled above one can see a difference between ideal and viscous solutions ,  shown in Figs.~\ref{fig_visc_rad} through the 
temperature dependence $T=\hat{T}/t$ at certain positions. 
The viscous effect is maximal at early times, while the viscous and ideal curves meet near freezeout.
As expected, the viscous effects are  more noticeable at the fireball edge, compare the $r=1$ and the $r=3$ plots. The main conclusion of this section is that small viscosity of the sQGP provides only modest corrections to
 the radial flow, even for the pA and pp cases.

\begin{figure}[t]
\begin{center}
\includegraphics[width=6cm]{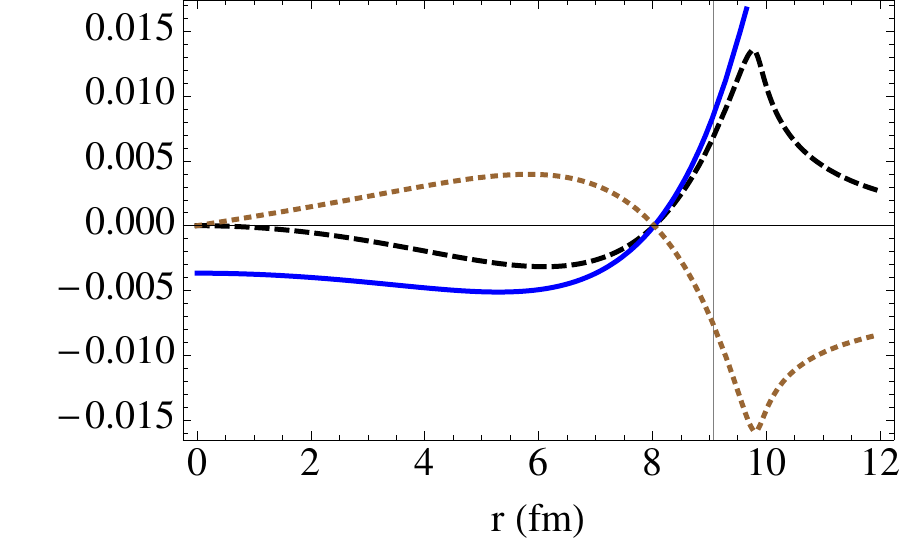}
\includegraphics[width=6cm]{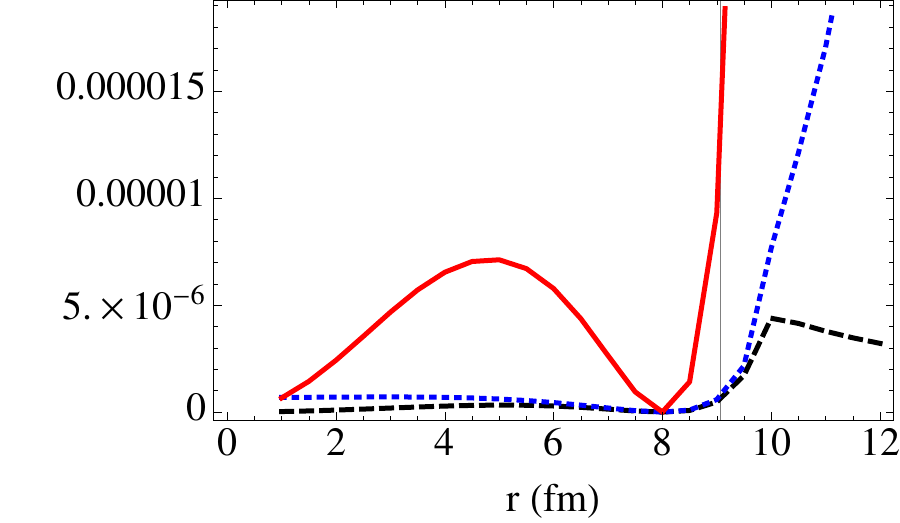}
\caption{(color online) Radial dependence of the first order viscous term (a) and the second order one (b) for central AA collisions.
 In (a) the black dashed,blue solid, and brown dotted lines show 00,11,01 components of $\sigma_{\mu\nu}$, respectively.
In (b) we show 00,11,22 components of (\ref{eqn_sigmasigma}) by black dashed, blue  dotted and red solid lines.
   }
\label{fig_sigma}
\end{center}
\end{figure}

   Another sourse of viscous corrections comes from modifications of the particle distributions induced by
   gradients of the flow. Those should be proportional to tensor of flow derivatives at the freezeout surface
   \be \delta f(x,p) \sim f(x,p) p^\mu p^\nu u_{\mu; \nu} \ee
   where semicolon as usual stands for covariant derivative. The coefficient is to be determined from 
   the fact that this correction is the one inducing the viscosity part of the stress tensor. 
Looking at the space-time dependence of the (symmetrized) tensor of flow covariant derivatives
\be  \sigma_{\mu\nu}= u_{<\mu; \nu>}\ee
we found rather curious behavior produced by Gubser's flow. In
 Fig.\ref{fig_sigma} we display several components of this tensor, and one can see that some of them change sign and magnitude
 at $r\approx 10 \, fm$, which is on the r.h.s. or space like part of the freezeout surface in AA collisions. (The ``corner" in this case is at $r\approx 9.1 \,fm$.) We think that this behavior is in fact an artifact of the Gubser solution caused by slow (power-like)
 decrease of the density at large distance. This tails of the matter distribution serve in fact as an``atmosphere"
 around the fireball, in which some fraction of expanding matter get accelerated inwards.  We checked that such behavior 
 is not observed for exponentially decaying tails, as is the case for real nuclei. Our conclusion then is that
 one should not use Gubser solution outside of the fireball ``rim", in our case for $r>9.1\, fm$. Fortunately,
 with realistic nuclear shapes that part of the surface contribute only very small -- few percents -- contribution to particle
 spectra and can therefore be neglected.

     Let us now start the discussion of the second and higher order gradients.
In general, those can be treated phenomenologically: one can write down a complete set of all
possible forms for the stress tensor of the given order, with some coefficients to be determined empirically.
The corresponding contribution to the stress tensor looks like
  \be 
  \pi^{\mu\nu} =\sum_n c_n  P_n^{ \mu\nu} (T,u^\alpha) \sim \sum_n  c_n   \left({1 \over TR}\right)^n
  \ee
  with some coefficients $c_n$ and certain kinematical structures with $i$ derivatives $P_n^{ \mu\nu} $. 
  Their order of magnitude is given by
  the pertinent powers of the hydro parameter $1/TR$, or multiplcity.  
Unfortunately, even for the second gradients there are way too many terms for that to be a practical program . 

For conformal fluids  the number of the second order terms is more manageable and using the AdS/CFT one can obtain the
value of the coefficients  (for review see \cite{Romatschke:2009im} ). Using such as a guide, one can estimate the magnitude of the terms neglected in the Navier-Stokes approximation.
Furthermore, for Gubser flow we find that the rotational  (antisymmetric) combination of the covariant derivatives
$\omega_{\alpha,\beta}=u_{[\alpha; \beta]}=0$, which eliminates two more terms. The term which is the easiest to estimate
is the symmetrized convolution of two first order term 
\be  \label{eqn_sigmasigma}
\pi_{\mu\nu}^{(2)}= - {\lambda_1\over 2} \sigma_{<\mu \lambda} \sigma^\lambda_{\nu>}  
\ee
where angular bracket stands for symmetrization of $\mu\nu$. The AdS/CFT value for the coefficient is $\lambda_1=\eta/(2\pi T)$.

Radial dependence of this term at the freezeout surface for AA collision is shown in  Fig.\ref{fig_sigma} (b).
It is reasonably small and constant, except strong growth ``beyond the rim" of the fireball. As we already noted above, 
this is the artifact of the Gubser solution, which should be ignored.

\subsection{The radial expansion and the Israel-Stuart second-order hydrodynamics}

    Using the lowest order hydrodynamics equations one can trade the spatial derivatives by the time ones,
and subsequently promote the ``static" gradient tensor $\sigma^{\mu\nu}$ to ``dynamical" stress $\pi^{\mu\nu}$,
with its own equaltion of motion. One may wander how these equations behave in the Gubser setting.

Since the first version of this paper was posted, this was done in \cite{Marrochio:2013wla}, which we follow in this section.
 The main purpose of this paper has been methodical, to check 
their previously developed MUSIC hydro solver against the analytically solvable examples.
(The solutions  discussed  were not intended to correspond to any particular physical settings.)

 The IS equations to be solved have in this case the form
\be {\hat T' (\rho)\over \hat T(\rho)}+{2\over 3} tanh(\rho) = {1 \over 3} \pi(\rho) tanh(\rho) \ee
\be c {\eta \over s}[  \pi'(\rho)+ {4 \over 3} \pi(\rho)^2 tanh(\rho)]+ \pi(\rho)\hat{T}(\rho)={4 \over 3}  {\eta \over s}  tanh(\rho) \ee
where  a prime denotes the derivative over the ``time" $\rho$,  and \be \hat T= T\tau  , \,\,\,\,\pi(\rho)=\hat\pi^\xi_\xi{1 \over \hat{T}\hat{s}}\ee  

Note that at $\rho\rightarrow \pm \infty$ the dimensionless temperature $\hat{T}$ vanishes as certain negative power of $cosh\rho$,
and therefore the second eqn decouples from the first. Furthermore, putting to zero the derivative, one find constant fixed point solution
$\pi=1/\sqrt{c}$, to which any solution should tend in the   $\rho\rightarrow \pm \infty$ limit. This feature is 
very unusual, in variance with Navier-Stokes and generic dissipative
equations, which only regulate solutions at positive time infinity, generating
singular or indefinitely growing solutions toward the past  $ \rho\rightarrow - \infty$. In this sense, there exists  clear advantage of the IS equations over the NS ones: but we don't think
 this  improvement reflects actual physics. 

\begin{figure}[t]
\begin{center}
\includegraphics[width=7cm]{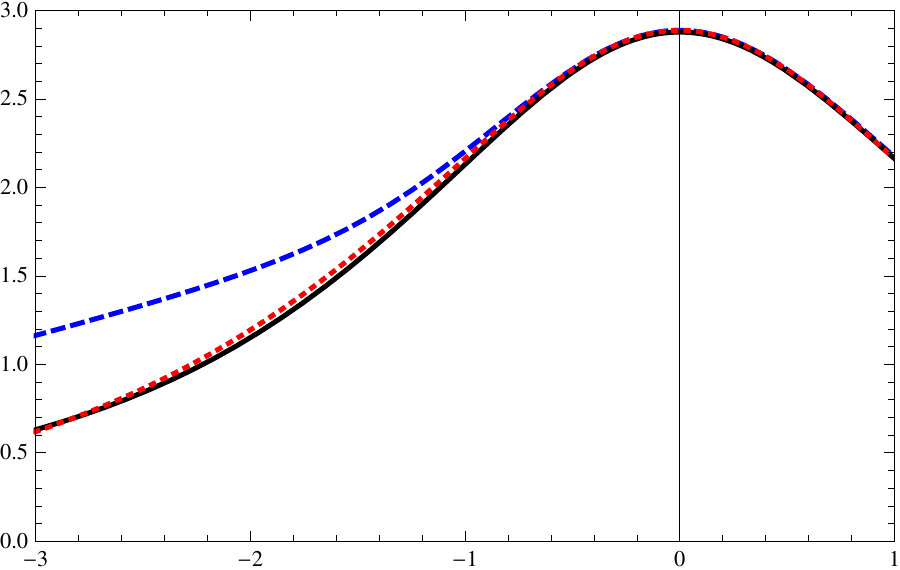}
\includegraphics[width=7cm]{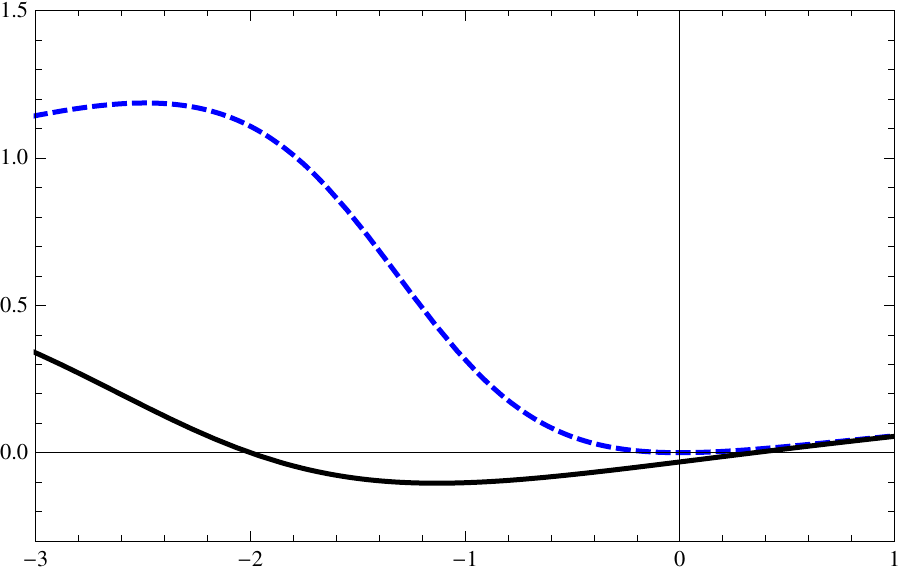}
\caption{(color online) (a) The dimensionless conformal temperature $\hat{T}$ and (b) the dimensionless
conformal stress $\pi$ as a function of ``time" $\rho$. The parameters correspond to $q=1 \, fm^{-1}, \eta/s=.2,c=5$ and multiplicity corresponding to the highest multiplicity bin of pA in CMS
experiment.
 The red dotted line in (a) is the ideal hydro Gubser solution $\hat{T}_0/cosh^{2/3}\rho$. In both plots
the blue dashed lines are for a ``realistic" solution with $\pi(0)=0$ near freezeout, while the black solid lines are for ``nice" solution with zero anisotropic stress at
the initiation time,
$\pi(-2)=0$. 
   }
\label{fig_IS}
\end{center}
\end{figure}

   The negative of
  the Israel-Stuart version of hydrodynamics, is that selecting the initial conditions for $\pi(\rho)$ is a nontrivial
task. In principle, some theory of pre-equilibrium conditions -- e.g. the AdS/CFT or color glass condensate (CGC) model -- should provide it.
For lack of knowledge about the initial value of the anisotropic part of the pressure tensor $\pi^{\mu\nu}$ 
 practitioners often
select  $\pi(\tau_i)=0$ at the initiation time, and then carry it on from the equation, till freezeout. 
In Fig.\ref{fig_IS} such a solution to Israel-Stuart equations given above is shown by the black solid lines. This solution is indeed more than satisfactory, in the sense that  
the temperature is very close to the ideal case (red dotted line), and $\pi$ remains small. 

This however is opposite to 
general expectations for the real QCD setting, in which the coupling constant runs from small to large as a function of time.
Because of that, the $\eta/s,c$ are not in fact constant but run, toward the most ideal fluid reached near $T_c$,  at the end of the QGP era. 
Therefore one expects the
 non-equilibrium effects -- in particular described by $\pi$ -- to monotonously decrease from the initial to the final state, as close to equilibrium as possible.
 We therefore suggest another possible solution, with  $\pi(\rho)$ set to  be zero at the end of the expansion, 
at the freezeout. This solution is shown in   Fig.\ref{fig_IS}  by the blue dashed line: it indeed shows a monotonous decrease of  $\pi(\rho)$ in the range of interest,
$\rho=-2..0$. While this scenario
it is not as nice as the previous one -- the anisotropic pressure is not small at the initial time $\pi(-2) \sim 1$ and in the temperature  deviations from 
the ideal solution are well seen -- perhaps it is closer to reality.  

In summary, while IS approach has advantages such as regular behavior of the solutions at both time infinities,
in practice it allows wide range of solutions in between, depending on the required initial conditions for the viscous
tensor. There is no real argument explaining why this version can be better than the first order NS in cases when viscous corrections get noticeable,
as there is no estimate of the terms neglected.

\subsection{Higher gradients and Lublinsky-Shuryak re-summation} \label{sec_higher_grad}

  The Navier-Stokes and Israel-Stuart approximations used so far only includes the first and the second order terms in the gradient expansion. What about high orders?

  The expansion coefficients may be obtained from
  AdS/CFT, an indispensable tool.
For small (linearized) perturbations -- sounds --
  the correlators of the two stress tensors was calculated to higher orders in frequency and wave vector
  $\omega,k$, extending the original  viscosity prediction $\eta/s=1/4\pi$  of Son et al to about a dozen further coefficients. 
  
  Can one  re-sum the higher gradient terms? While hydrodynamics is more than two centuries old, it seems that
  the first attempt of the kind has been suggested by Lublinsky and Shuryak (LS) \cite{Lublinsky:2009kv}.
 An approximate PADE-like re-summation of the higher order terms 
 results from the alternating signs of the series and coefficients of the order 1, which calls for
approximate re-summation {\em a la}  geometrical series\footnote{The reader may ask why not other series, such as e.g.
leading to $e^{-x}$. While there is not enough known higher order terms to tell the difference, large $x$ behavior
of the geometric series seems to us more appropriate.} 
\be 1-x+x^2 +\ldots \rightarrow {1 \over 1+x} \label{expansion} \ee
which keeps the quantity positive and regular even for $x>1$. 
The suggested recipe is to substitute
the Navier-Stokes viscosity constant
 by an effective one, which is in frequency-momentum dependent and
reads
\be  
\eta_{LS2}(\omega,k) ={ \eta_{NS} \over 1- \eta_{2,0} k^2/(2\pi T)^2 -i \omega \eta_{0,1}/(2\pi T)} 
\label{MISHA}
\ee
while (\ref{MISHA}) involves only two dimensionless coefficients, whose values for AdS/CFT are 
\be \eta_{2,0} =-{1\over 2} \,\,  \eta_{0,1}=2-{\rm ln2}=1.30\ee
 it actually approximately reproduces about a dozen of known terms.
Note that re-summation into the denominator suggests a $reduction$ of the viscous effect as gradient grows.
It may look counterintuitive: note however that viscosity is a coefficient of a term in hydro equations with at list second order of $k$:
so this reduction only makes such terms finite, not zero.

 Recently one of us has studied the ``strong shock wave" problem
 \cite{Shuryak:2012sf} in the AdS/CFT setting, solved from the first principles (Einstein equations) and comparing to the LS re-summation. While  this problem is far from sound and is a generic ``hydro-at-its-edge" type,  with large gradients
 without any small parameters, 
  deviations between the NS and the exact (variational) solution
 of the corresponding Einstein equations were  found to be on the level of few percents only. 
 Studies of time-dependent collisions in bulk AdS/CFT have found that the first-principle solution
 approaches the NS solution early on and quite accurately, at the time when the higher gradients by themselves are not small,
see e.g. \cite{JANIKHIGHER}.
 
    Let us now check how does it work in the case of Gubser solution.
 Changing $k^2,\omega$ into derivatives
 \ba -k^2/q^2 &\rightarrow & ({\partial \over \partial r})^2+{1 \over r} {\partial \over \partial r} \nonumber \\
 i\omega/q &\rightarrow & {\partial \over \partial t}
  \ea
  makes the re-summed factor (with the denominator)  an integral operator, which can be used
  not only for plane waves of the sound but for any function
   of the coordinates $f(t,r)$. The inverse ``LS operator" acting on a function $f$ is defined as
 \ba 
 {\bf O}_{LS}^{-1}(f)=&&
  1+{q^2 \over 2(2\pi T)^2}\left({\partial^2 f\over \partial r^2}+{1 \over r} {\partial f \over \partial r}\right){1\over f} \nonumber \\
 && +\,\,(2-{\rm ln2})\, {q  \over 2\pi T}{\partial f \over \partial t} {1\over f}
\label{eq_LS}\ea 
Schematically the resummed hydro equations look as 
\be (Euler)= \eta  {\bf O}_{LS} (Navier-Stokes) \ee
where ${\bf O}_{LS}$ is an integral operator. 
However, one can act with its inverse on the hydrodynamical equation as a whole, 
acting on the Euler part but canceling it in the viscous term
\be  {\bf O}_{LS}^{-1} (Euler) = \eta (Navier-Stokes) \ee
  These are the equations of the LS hydrodynamics. 
Obviously they have two extra derivatives and thus need more initial conditions for solution.

   Instead of solving these equations, we will simply check the magnitude of the corrections appearing in the l.h.s 
   due to the action by the LS differential operator
 on the (ideal Gubser) solution used as a zeroth-order starting point. 
 As one can see, large systems have a small $q/T\sim 1/RT$ parameter and so these corrections 
are $parametrically$ small.  
The issue is what happens ``on the hydro edge", when the
corrections have no formal small parameter.

\begin{figure}[t]
\begin{center}
\includegraphics[width=5cm]{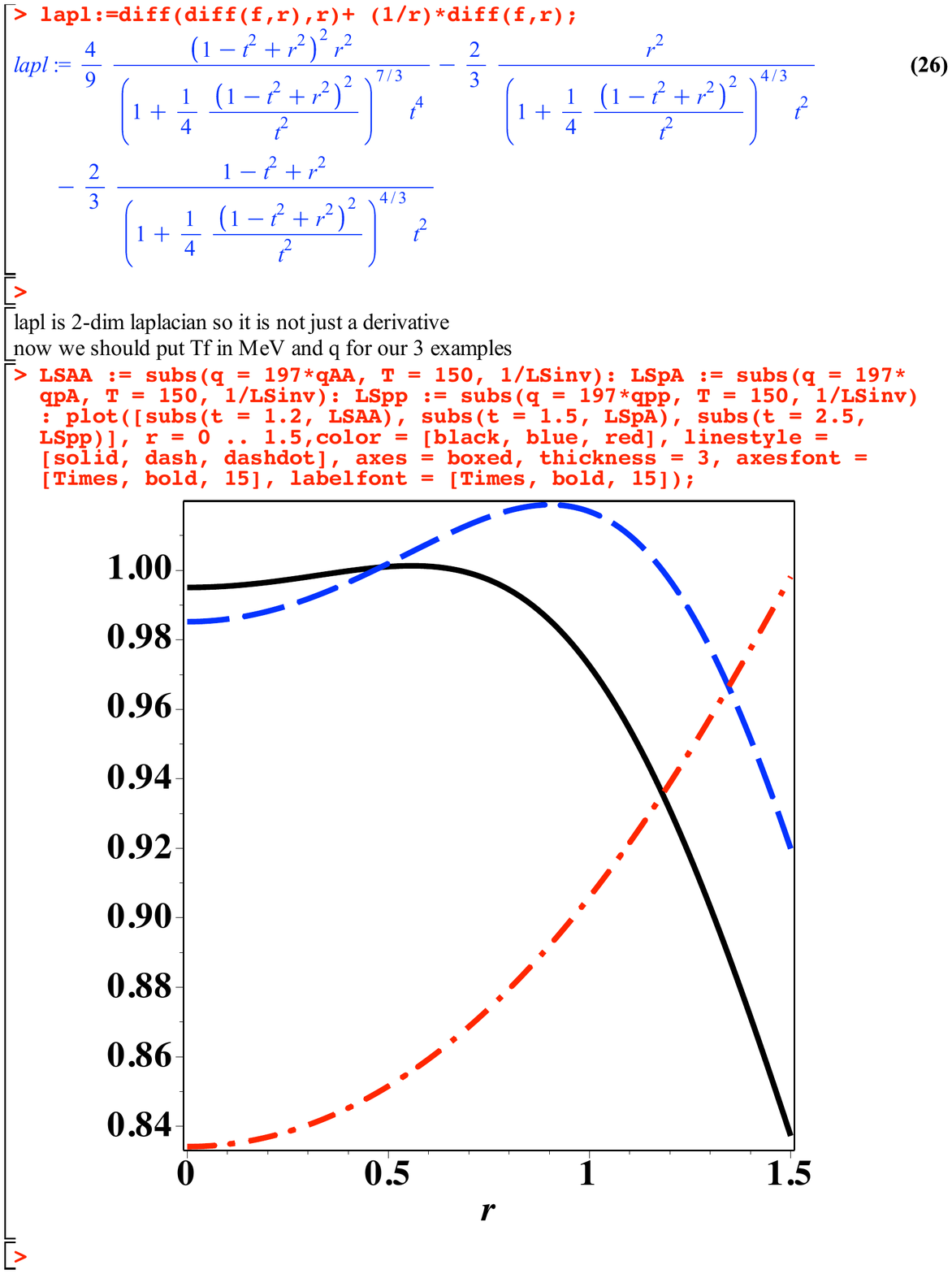}
\caption{(color online) The action of the LS operator $O_{LS}$ (\ref{eq_LS}) on the zeroth order (non-viscous) temperature profile,
the first term of (\ref{back_T}).
The three  lines correspond to  AA (black) solid, pA (blue) dashed
 and pp (red) dash-dot. 
   }
\label{fig_LS_corr}
\end{center}
\end{figure}

In Fig.\ref{fig_LS_corr} we show the (inverse) action of (\ref{eq_LS}) on the zeroth other temperature profile of the Gubser flow
as a function of $r$. We have used the freeze-out temperature $T_f=150\, MeV$ and the indicated
respective freeze-out times for pp, pA and AA.
The higher gradient corrections for AA and pA are inside the few percent range from 1,
while in the pp case the correction is larger, yet still in the 15 percent range. We thus conclude, that if the LS 
resummation represents the role of the higher gradients, the overall corrections remain manageable, although it does grow from 
AA to pA to  pp cases.

\section{Higher angular harmonics}
\subsection{Acoustic damping}
   There is a qualitative difference between the radial flow we had discussed so far, and higher angular harmonics.
   While the former  monotonously grows with time, driven by sign-constant pressure gradient, the latter are
 a (damped) oscillators. The signal observed depend on the viscous damping factor as well as
on the particular phase in which the oscillator finds itself at the freezeout time. We will discuss those effects subsequently.

The effects of viscosity  damps  the higher angular flow moments stronger. The so called ``acoustic damping" formula
 was suggested by Staig
and Shuryak~\cite{Staig:2010pn} . Wave amplitude reaction is given by 
  \be
P_k={\delta T_{\mu\nu} (t,k) \over   \delta T_{\mu\nu} (0,k)
} = {\rm exp}\left(-{2 \over 3} {\eta \over s}
{k^2 t \over  T } \right) \label{eqn_visc_filter} \ee 
Since the scaling of the freeze out time is linear in $R$ or $t_f \sim R$, and the wave vector $k$ corresponds to
the fireball circumference which is $m$ times the wavelength
\be 2 \pi R =m {2\pi \over k} \ee 
 the expression (\ref{eqn_visc_filter}) yields
 
\be {v_m\over \epsilon_m}\sim {\rm exp}\left[ - m^2 {4 \over 3} \left({\eta \over s}\right)\left({1 \over TR}\right) \right]  \label{eqn_acoustic}  \ee
Note that the exponent contains the product of two small factors,  $\eta/s$ and $1/TR$, as discussed in the introduction.
 Note further that  the harmonics number is squared.
 For central PbPb LHC collisions with
 \be {1 \over TR} = {\cal O}(1/10) \ee  
 its product of $\eta/s$ is $O(10^{-2})$. So one can immediately see from this expression
why  harmonics  up to  $m=O(10)$ can be observed.

Proceeding to snapper systems in the spirit of our thought experiment 0, by keeping a similar initial temperature $T_i\sim 400 \, MeV \sim 1/(0.5\, fm)$
but a smaller size $R$,  results in a macro-to-micro parameter that is no longer small, or 
$1/TR\sim 0.5,1$, respectively. 
For a usual liquid/gas, with $\eta/s>1$, there would not be any small parameter left and one would have to conclude
that hydrodynamics is inapplicable for such a small system. However, since the quark-gluon plasma is an exceptionally
good liquid with a very  small  $\eta/s$, one can still observe harmonics up to  $m=O(\sqrt{10})\sim 3$.
However, if $TR=const$, along  the line of the thought experiment 1, there is no difference in the damping. 

   Extensive comparison of this expression with the AA data, from central to peripheral, has been
recently done in Ref.  \cite{Lacey:2013is} .
Both issues -- the $m^2$ and $1/R$ dependences of the $log(v_m/\epsilon_m)$ --
are very well reproduced. It works all the way to rather peripheral AA collisions with $R\sim 1 \, fm$
and multiplicities comparable to those in the highest pA binds. Thus the acoustic damping provides solid hydro-based systematics
of the harmonic strength, to which new pA  and pp data  should be compared.
  
\subsection{Angular harmonics of  Gubser  flow \label{sec_harm}}

   Unfortunately,  the acoustic damping formula does not include the oscillatory prefactors.
 (As emphasized in Ref.  \cite{Staig:2011wj}, those should lead to secondary peaks in power spectrum of fluctuations at high $m$ similar 
 to  those in cosmological perturbations. Those are however not yet observed.) 
   
Since we are actually interested in
not so large $m=2,3$,  we return to
Gubser's flow and consider its angular perturbations. 
Those has been developed in \cite{Gubser:2010ui,Staig:2011wj}. 
In the former paper Gubser and Yarom re-derived the radial solution
 by going into the co-moving frame
via a coordinate transformation from the $\tau,r$ to a new set $\rho,\theta$ given by:
\begin{eqnarray}
\sinh{\rho} & = & -\frac{1-\tau^2+r^2}{2\tau}\label{rho_coord}\\
\tan{\theta} & = &
\frac{2r}{1+\tau^2-r^2}\label{theta_coord}
\end{eqnarray}
In the new coordinates the rescaled metric reads:
\begin{eqnarray}
d\hat{s}^2 & = &-d\rho^2 + \cosh^2{\rho}\left(d\theta^2 +
\sin^2{\theta}d\phi^2\right)+d\eta^2 \nonumber
\end{eqnarray}
and we will use $\rho$ as the ``new time" coordinate and $\theta$ as a
new ``space" coordinate. In the new coordinates the fluid is at
rest, so the velocity field has only nonzero $u_{\rho}$. 
The  temperature  is now dependent only on the
new time $\rho$. For nonzero viscosity the solution is
\begin{eqnarray}
&&\hat{T} =\frac{\hat{T}_0}{(\cosh{\rho})^{2/3}} +\frac{H_0
\sinh^3{\rho}}{9 (\cosh{\rho})^{2/3}} \,\nonumber\\
&&\,\,\,\,\times\,\,_2F_1\left(\frac{3}{2},\frac{7}{6};\frac{5}{2},-\sinh^2{\rho}\right)
\label{back_T}
\end{eqnarray}
with $\hat{T}=\tau f_*^{1/4}T$ and $f_*=\epsilon/T^4=11$ as in \cite{Gubser:2010ze}.

Small  perturbations to GubserÕs flow obey linearized equations which have also been derived in \cite{Gubser:2010ui}. We start with  the zero viscosity case, so that the background temperature (now to be called $T_0$) will be given by just the first term in (\ref{back_T}). The perturbations over the previous solution are defined by
\begin{eqnarray}
\hat{T} & = &  \hat{T}_0(1+\delta)\label{Tpertb}\\
u_{\mu} & = & u_{0 \,\mu} + u_{1\mu}\label{upert}
\end{eqnarray}
with
\begin{eqnarray}
\hat{u}_{0 \,\mu} & = & (-1,0,0,0)\\
\hat{u}_{1\mu} & = & (0,u_{\theta}(\rho,\theta,\phi),u_{\phi}(\rho,\theta,\phi),0)\\
\delta & = & \delta(\rho,\theta,\phi)
\end{eqnarray}
The exact solution can be found by using the 
separation of variables 
$\delta(\rho,\theta,\phi)=R(\rho)\Theta(\theta)\Phi(\theta)$. In the non-viscous case, that we are now discussing, each of the three equations
\begin{eqnarray}
&&RÕÕ(\rho)+\frac{4}{3}\tanh{\rho}RÕ(\rho)+\frac{\lambda}{3\cosh^2{\rho}}R(\rho)=0\nonumber\\
&&\ThetaÕÕ(\theta)+\frac{1}{\tan{\theta}}\ThetaÕ(\theta)+\left(\lambda-\frac{m^2}{\sin^2{\theta}}\right)\Theta(\theta)=0\nonumber\\
&&\PhiÕ(\phi)+m^2\Phi(\phi)=0
\end{eqnarray}
are analytically solvable, with the results discussed in  \cite{Staig:2011wj}. 
%
 The parts  of the solution
depending on $\theta$ and $\phi$ can be combined in order to form
spherical harmonics $Y_{lm}(\theta, \phi)$, such that
$\delta(\rho,\theta,\phi)\propto R_l(\rho)Y_{lm}(\theta,\phi)$.

The basic equations for the $\rho$-dependent part of the perturbation, now with viscosity terms, can be
written as a system of coupled first-order equations
 \cite{Gubser:2010ui}. We are  assuming
rapidity independence, thus the system of equations
$(107)$,$(108)$ and $(109)$, from the referred paper, becomes two
coupled equations, for  (the $\rho$-dependent part of) the
temperature and velocity perturbations
\begin{eqnarray}
\frac{d\vec{w}}{d\rho}= -\Gamma \vec{w} \,\, ,\,\,\,\,\,\, \vec{w} & = &\left(%
\begin{array}{c}
  \delta_v \\
  v_v \\
\end{array}%
\right) \label{visMeq}
\end{eqnarray}
where the index v stands for viscous and the matrix components
are, 
\ba
&&\Gamma_{11} =  \frac{H_0 \tanh^2\rho }{ 3\hat{T}_b } \nonumber\\
&&\Gamma_{12}  =  \frac{l(l+1)}{3\hat{T}_b \cosh^2{\rho}}
\left(H_0\tanh{\rho}-\hat{T}_b\right) \nonumber\\
&&\Gamma_{21}  =  \frac{2H_0 \tanh{\rho}}{H_0\tanh{\rho}-2\hat{T}_b}+1 \label{visMatrix} \nonumber\\
&&\Gamma_{22} =  ( 8\hat{T}_b^2\tanh{\rho} \nonumber\\
&&+ H_0\hat{T}_b
\left( \frac{-4( 3l( l+1 )- 10 )) }{ \cosh^2{\rho}}-16 \right) \nonumber \\ 
&&+6H_0^2\tanh^3{\rho} )/( 6\hat{T}_b \left( H_0\tanh{\rho} -
2\hat{T}_b \right) ) \nonumber\\
\ea

\begin{figure*}[t!]
\begin{center}
\includegraphics[width=7cm]{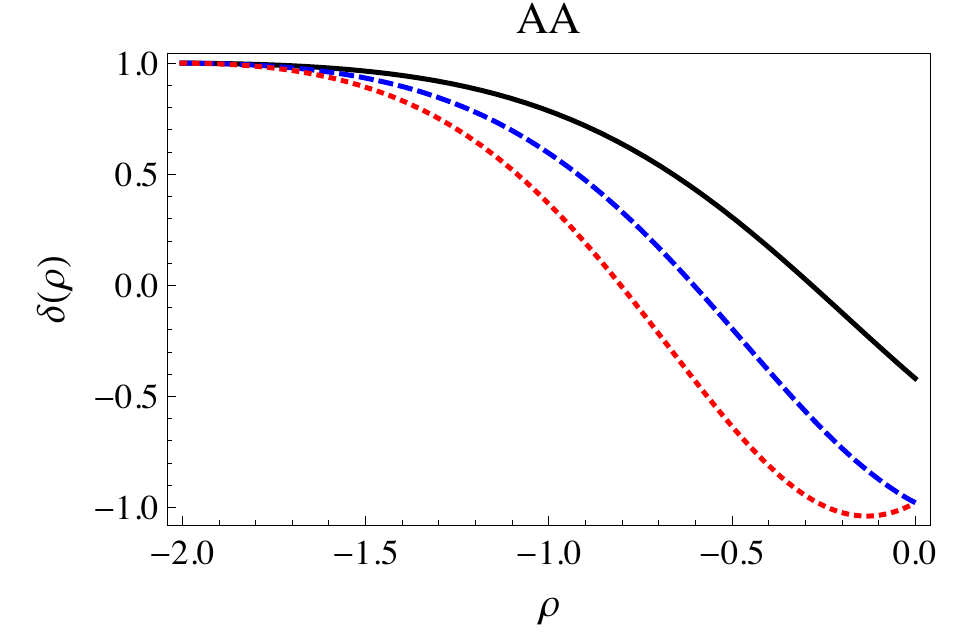}\includegraphics[width=7cm]{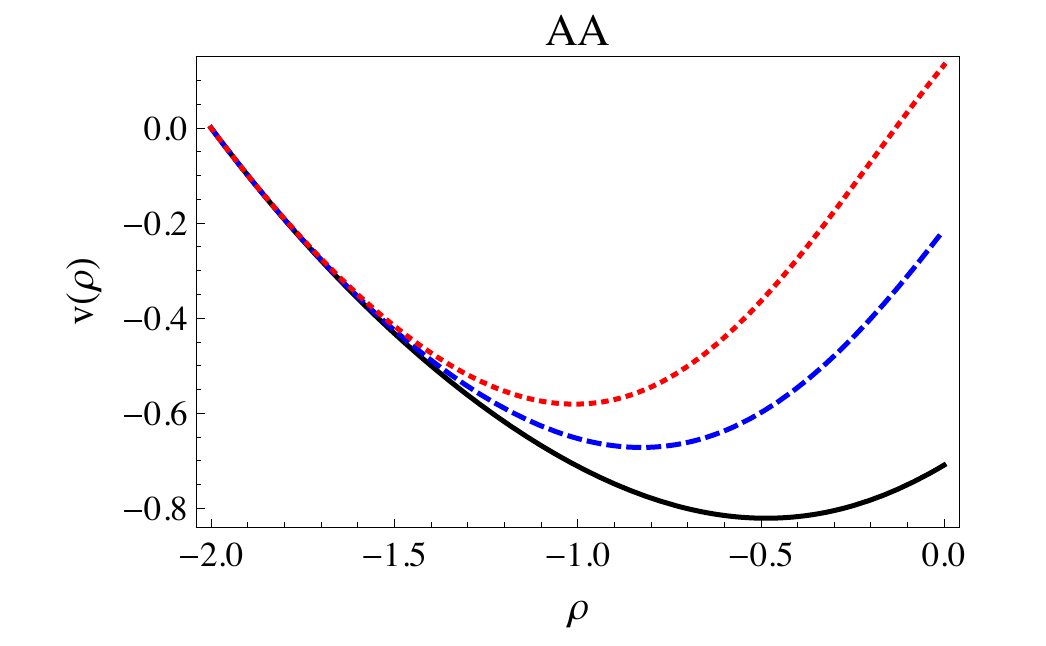}\\
\includegraphics[width=7cm]{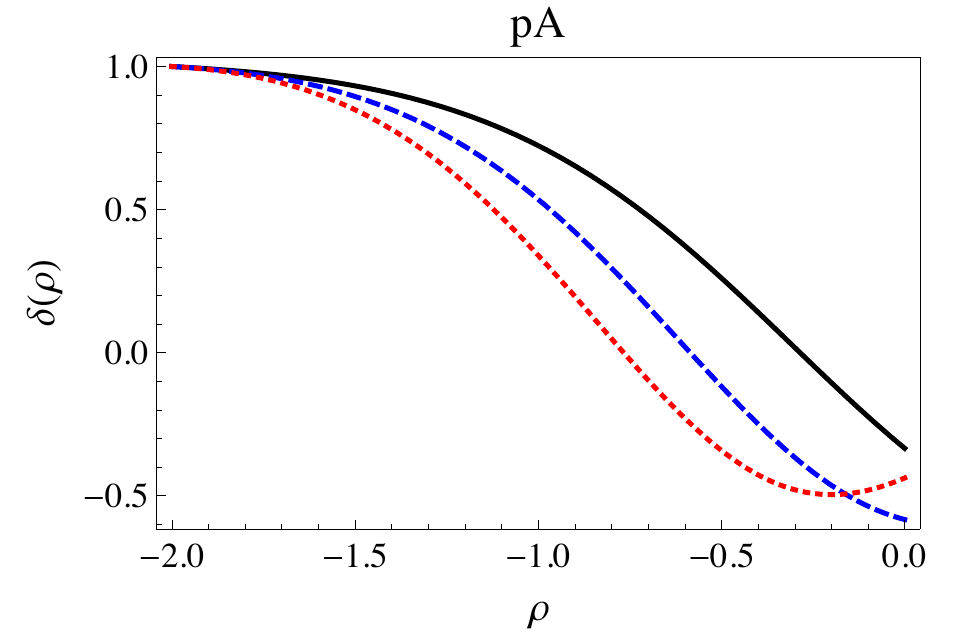}\includegraphics[width=7cm]{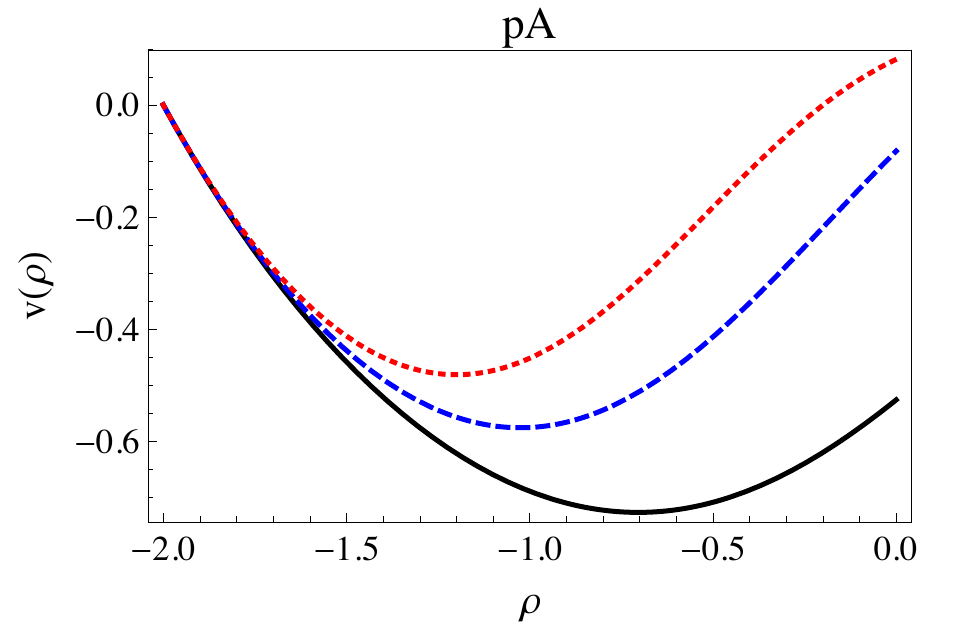}\\
\includegraphics[width=7cm]{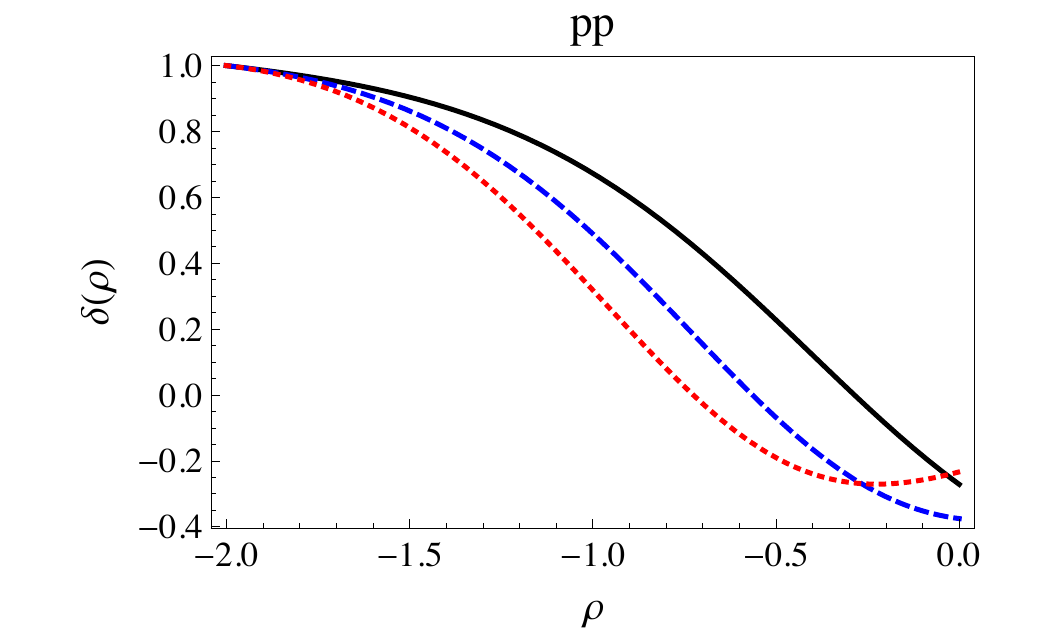}\includegraphics[width=7cm]{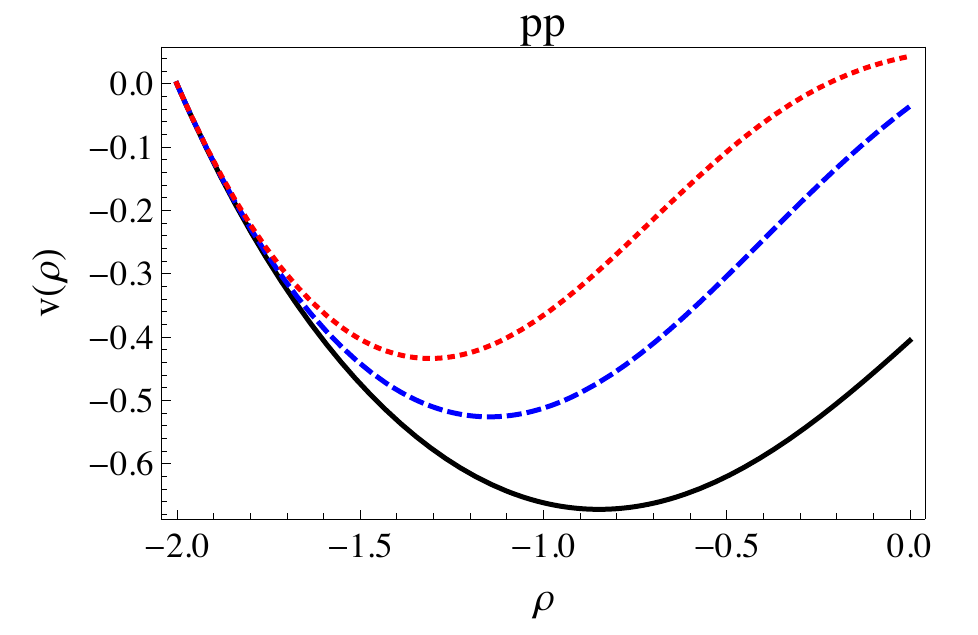}\\
\caption{(color online) The dimensional less temperature perturbation $\delta_l(\rho)$ 
 and velocity  $v_l(\rho)$, for $l=2,3,4$ shown by (black) solid, (blue) dashed and (red) dotted curves, respectively.
 Three sets of calculations corresponds to AA, pA and pp collisions. }
\label{fig_harm_gubser}
\end{center}
\end{figure*}

Before we display the solutions, we need to translate our space-time plot into the $\rho-\theta$ coordinates.
The initiation surface $t=t_i$ are $not$ the $\rho=const$ surfaces. The freezeout ones  also do not correspond to
fixed $\rho$ because the temperature is $T=\hat{T}(\rho)/t(\rho,\theta)$. 
So, in both cases one has to decide which 
points on the initiation and final surfaces are most important. The thin solid lines in Fig.\ref{fig_tr_ideal}
approximately 
represent the initial $\rho_i$ and the final $\rho_f$ values for all three systems. Therefore, we will solve
the equations between those two surfaces.

In Fig.\ref{fig_harm_gubser} we show the solution of the $\rho$ evolution of the two variables, the temperature perturbation 
and velocity $\delta_l(\rho),v_l(\rho)$. As one can see, all of them start at $\rho_0=-2$ from the same $\delta_l=1$
value. While the elliptic one $l=2$ (black solid curves) changes more slowly, higher harmonics oscillate more. 
We return to its discussion in section \ref{sec_pheno_harm}

\begin{figure}[t!]
\begin{center}
\includegraphics[width=5cm]{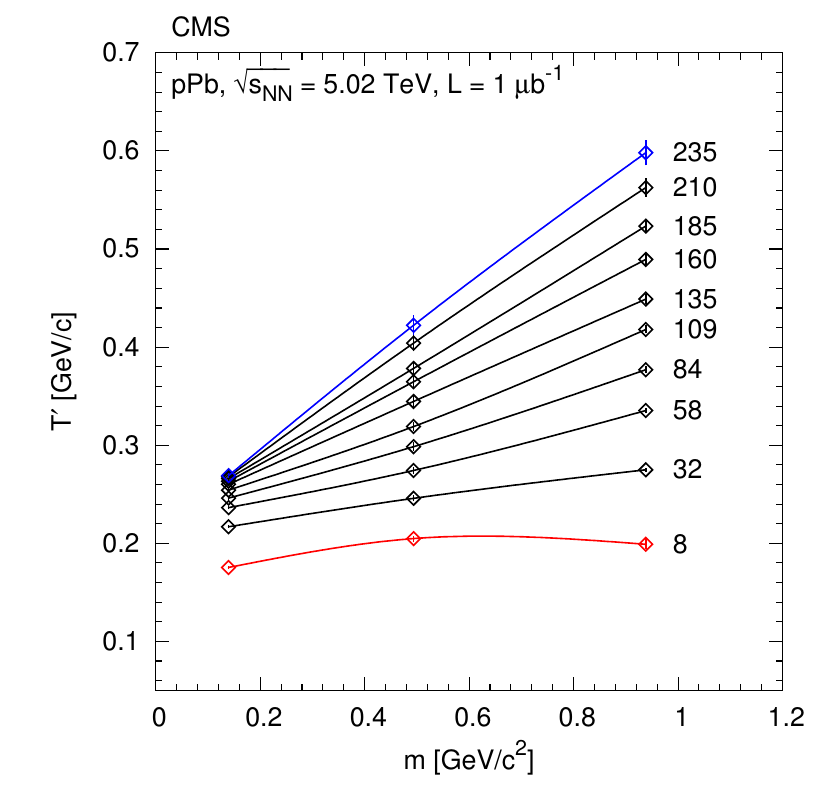}
\caption{(color online) The slopes of the $m_\perp$ distribution $T'$ (GeV) as a function of the particle mass, from   \cite{Chatrchyan:2013eya}. The numbers on the right are track multiplicity.}
\label{fig_slopesa}
\end{center}
\end{figure}

\begin{figure}[t!]
\begin{center}
\includegraphics[width=5cm]{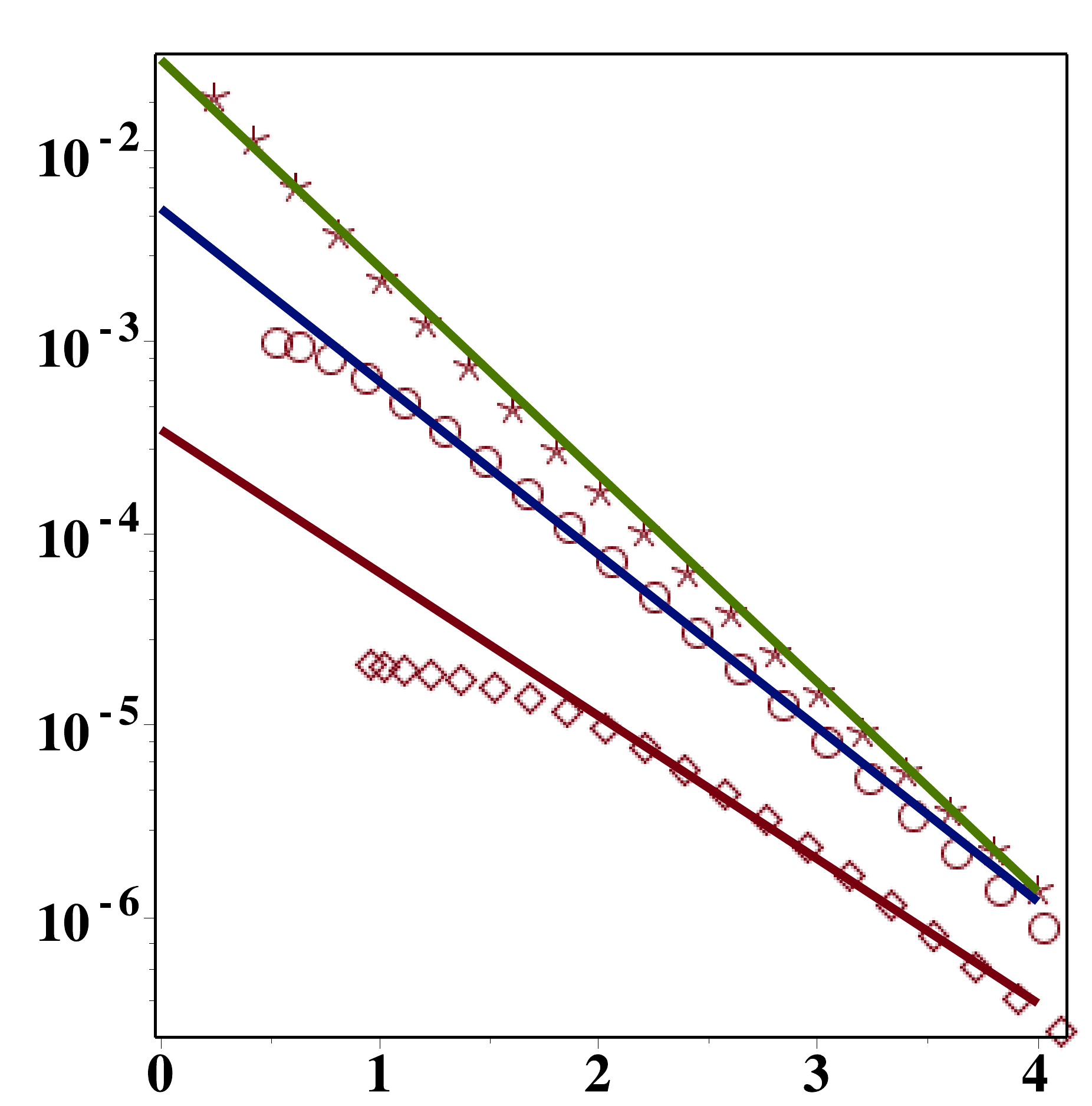}
\includegraphics[width=5cm]{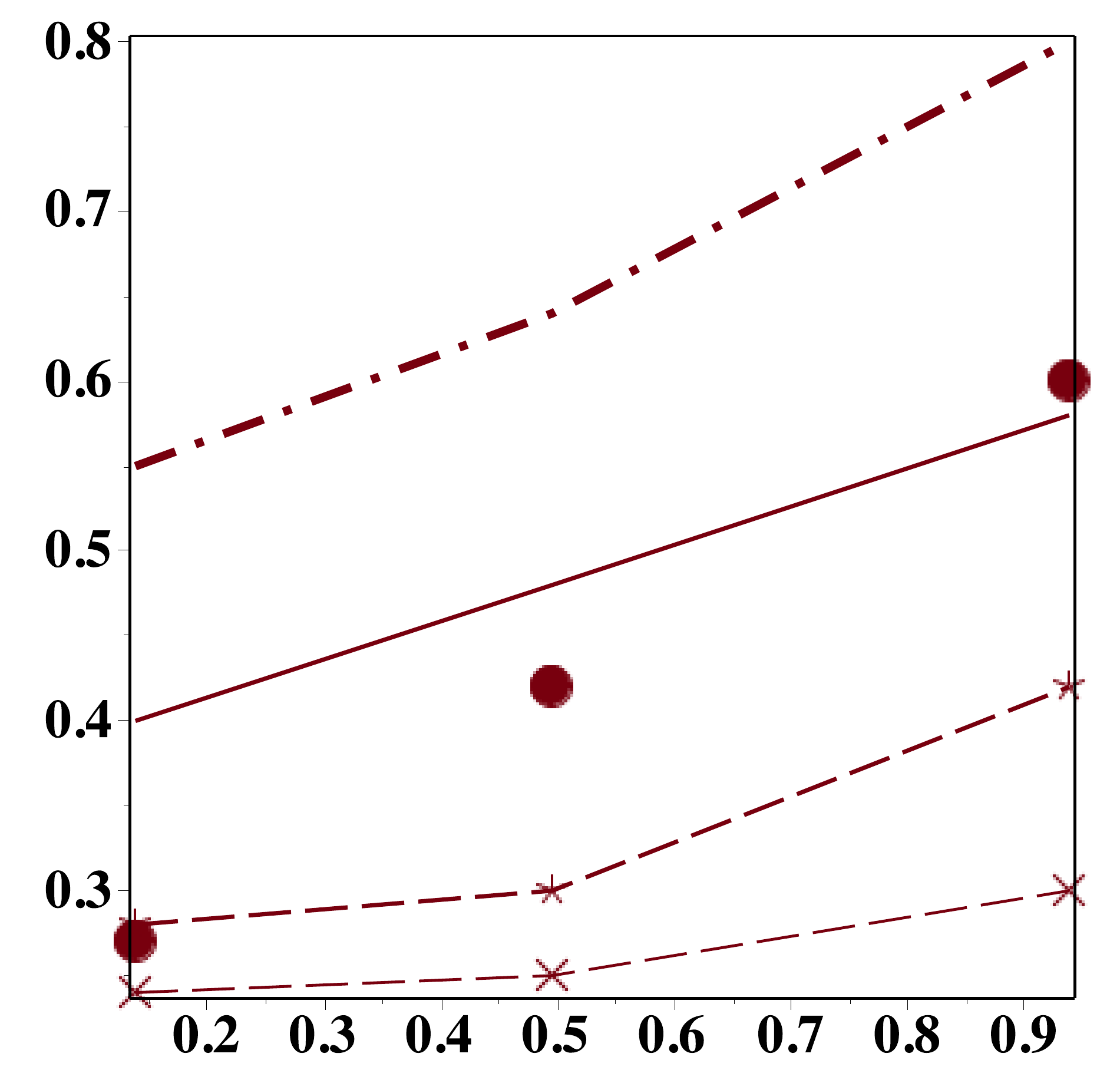}
\caption{(color online)  (a) A sample of spectra calculated for $\pi,K,p$, top-to-bottom, versus  $m_\perp$ (GeV), together with fitted exponents. (  b) Comparison of the experimental slopes $T'(m)$ vests the particle mass $m$ (GeV). The solid circles are from the highest multiplicity bin data of fig.(a), compared  to those of the theoretical models.  The solid and dashdotted lines are our calculations for freezeout temperatures $T_f=0.17,0.12 \, GeV$. Asterisks/dashed line are for Epos LHC, diagonal crosses /dashed line are for AMTP models.}
\label{fig_slopesb}
\end{center}
\end{figure}

\section{Phenomenology}
\subsection{The radial flow in spectra of identified secondaries}
   
        The main idea behind experimental signatures of the radial flow has been  used in \cite{Shuryak:1979ds}, it
 is based on the fact that collective flow manifests itself differently for secondaries of different mass.     The exponential thermal spectra of the near-massless
pion are simply blue-shifted by a factor, the exponent of the transverse flow rapidity
$ T'=T e^\kappa$.  However
 spectra of  massive particles -- such as  kaons, protons etc -- are modified in a more complex way.
Instead of discussing the shape of the spectra, let us focus on their high-momentum behavior 
 and the so called $m_\perp$ slopes: the particle spectra are fitted to the exponential  form (above certain $p_t$)
 \be {dN \over dy dp_\perp^2} = {dN \over dy dm_\perp^2} \sim exp( -{m_\perp \over T'} ) \ee
 in the transverse mass variable $m_\perp=\sqrt{m^2+p_\perp^2}$, typically above certain value of the $m_\perp$ (see examples below).
 It has been found in~\cite{Shuryak:1979ds} using 
the min.bias ISR pp data that the so called ``$m_\perp$ scaling" holds -- the slopes $T'$ are {\em the same} for $\pi,K,p$ independent on their mass $M$. 
This scaling (coming from the string fragmentation mechanism) implies that there was $no$ evidence for collective expansion
in min.bias. pp collisions at the ISR energies.
 
     Recent CMS pA data  \cite{Chatrchyan:2013eya} significantly increased the range of multiplicities,
  and now contain spectra of identified particles.    
As seen in the shown in Fig.\ref{fig_slopesa}, for small multiplicity bins (marked by $8$ and $32$ at the bottom) the same $m_\perp$ scaling holds, 34 years later and at beam energies
hundreds of times higher. However  for larger multiplicity bins the slopes grow with the particle mass  linearly. Qualitatively similar behavior has been previously seen in AGS/SPS/RHIC and
LHC AA data, and is widely recognized as {\em the signature} of the radial flow.      Furthermore, six months after the first version of this paper \cite{Shuryak:2013ke} made its main prediction -- that not only  the radial flow in pA and pp will be observed, but that its magnitude will even be $larger$ than in central AA collisions -- is confirmed. 
The highest multiplicity pA do have slopes exceeding even those in central PbPb LHC collisions,  the previous record-holding on the radial flow.

  In Fig.\ref{fig_slopesb}(a)  we show samples $m_\perp$ spectra calculated from Gubser radial flow. As for any axially symmetric case,
  one can perform the integrals over the spatial rapidity
 and azimuthal angle analytically, both producing Bessel functions,
 \ba  {dN \over dy dp_\perp^2}=   {g_{stat} \over 2\pi^2}   \int dr \tau( r) r  \\ \nonumber  
 [ m_\perp K_1({m_\perp cosh(\kappa) \over T_f})  I_0({p_\perp sinh(\kappa) \over T_f})  \\ \nonumber  
 -  p_\perp {d\tau \over dr}  K_0({m_\perp cosh(\kappa) \over T_f})  I_1({p_\perp sinh(\kappa) \over T_f})] 
            \ea
 In the remaining radial Cooper-Fry
 integral over the freezeout surface one should substitute proper time $\tau( r)$ and its derivative,
 as well as transverse rapidity $\kappa(\tau( r),r)$, defined via $\tanh(\kappa)=v_\perp$.
 The spectra are  fitted to exponential form at large $m_\perp$ (see  Fig.\ref{fig_slopesb}( a)) and    
finally in Fig.\ref{fig_slopesb}( b) we compare the  slopes $T'$ observed by the CMS (in the highest multiplicity bin) to theoretical 
results. 

 We start doing it by comparing to other models. We
 do not include the parton cascade models Hijing, as it has no flow by design and obviously fails in such a comparison.  The (latest version of the) hydrodynamical model  ``Epos LHC" \cite{Pierog:2013ria} predicts spectra with slopes shown by asterisks: as evident from Fig( b) it  misses the slope by a lot, for the protons by about factor 2. Even further from the data are the slopes calculated from the AMPT model \cite{AMTP} (diagonal crosses and dashed line).
 
 Upper two lines  in Fig.\ref{fig_slopesb}( b)  show our results, corresponding to two selected values of $T_f$, .12 and .17 GeV. The former is in the ballpark of the kinetic freezeout
 used for AA data: but as the figure ( b)  shows it overpredicts the radial flow for the pA case. The second value corresponds to the QCD critical temperature $T_c$:
 it is kind of the upper limit for $T_f$ since it is hard to imagine freezeout in the QGP phase. As seen from the figure, such value 
 produces reasonable amount for the collective radial flow as observed by the CMS. The same level of agreement holds not
 only in the highest multiplicity bin, but for most of them.  We thus conclude that in pA the chemical and kinetic freezeout coincide.

   Apart from the effective $m_\perp$ slopes $T'$ for each multiplicity bin and particle type, the paper  \cite{Chatrchyan:2013eya} also  
 gives the mean
   transverse momenta. Like slopes, they also display that radial flow in few highest multiplicity pA do exceed that in central AA. 
   Those data also agree reasonably well with our calculation.
   
    (The reader may wander why we don't compare the spectra themselves. Unfortunately we cannot do it now, neither in normalization more in shape
  because of significant ``feed-down"  from multiple resonance decays, strongly distorting  the small-$p_t$ region.
   Event generators like HIJING and AMPT  use ``afterburner" hadron cascade codes for that.)

\subsection{Higher harmonics \label{sec_pheno_harm}}
   The repeated motive of this paper is that the smaller systems should have stronger radial flow, as they
   evolve ``longer" (in proper units, not absolute ones) and the pressure gradient driving them never disappears. 
Higher harmonics are not driven permanently but are instead oscillating, plus damped by the viscosity. 
Since the only harmonics in the pA and pp observed so far are the elliptic $m=2$ and triangular $m=3$ ones,
and their  origins are quite different, we will discuss them subsequently, starting from qualitative expectations and then returning to hydro calculations.

Elliptic deformation $\epsilon_2^{AA}$ of the peripheral AA collisions is quite large, significantly larger than than those of the very central pA $\epsilon_2^{pA}$ of 
comparable multiplicity. 
However if those are evaluated -- e.g. in the Glauber model -- and divided out in the ratio $v_2/\epsilon_2$, the result should be about the same in both cases
for the same multiplicity in both
sets. 
This is e.g. seen from the acoustic damping  expression (\ref{eqn_acoustic} ): the same value of the multiplcity/entropy implies the same $TR$ and thus damping. 

Specially interesting case is dAu collisions, as in this case there are two collision centers and $\epsilon_2$ is factor 2 enhanced \cite{Bozek:2011if}, 
and $v_2$ is also a factor 2 higher \cite{Adare:2013piz}

The m=3 flow originates from fluctuations, not from a particular average shape. 
Therefore\footnote{This consideration has been proposed by D.Teaney in
the discussion.} , assuming we compare the same number of wounded nucleons and  multiplicity in central pA and peripheral AA,
we expect similar $\epsilon_3$ in both cases. Indeed, the magnitude of all $m>2$ deformations is $\epsilon_{m>2}\sim 1/\sqrt{N}$ 
where $N$ is the number of ``fluctuating clusters" -- wounded nucleons. Thus we expect $v_3$ for both cases be {\em the same},
even without the need to renormalize it by $\epsilon_3$. This  predictions is indeed fulfilled in the LHC data.

   Now we return to hydro results, discussed in section \ref{sec_harm}. As one can see from Fig.\ref{fig_harm_gubser}
 the time from initial $\rho \sim -2$ till freezeout  $\rho \sim 0$  is between a quarter and a half
 of the period of the oscillations. 
So the energy associated with the initial spatial deformation is transferred into kinetic energy of the flow, and start to come back
when the explosion ends.  The amplitude of the velocity at the r.h.s. of the plot is the largest for the m=2, and is smaller
for m=3,4. Smaller system do evolve a bit ``longer" which put their velocity amplitudes closer to zero.

Note that all of those start from the same deformation $\epsilon_m$, so what one reads from this plot is actually
proportional to $v_m/\epsilon_m$. In order to get absolutely normalized 
 $v_m/\epsilon_m$ one has to do integration over the the Cooper-Fry freezeout, as done in the previous section for radial
 flow. Since the latter includes rather lengthy calculations (see \cite{Staig:2011wj} for  details) we will not do it at this stage.

Assuming that the integrals produce the same factors for all cases,
we just read off the ratios of 
 $v_m/\epsilon_m$ from Fig.\ref{fig_harm_gubser} values of the flow at $\rho\sim 0$.
Since measurements are for two-particle correlation functions, we compare
the $squares$ of the flow harmonics
\ba 
&&({v_2^{AA} \over \epsilon_2^{AA}})^2: ({v_2^{pA} \over \epsilon_2^{pA}})^2: ({v_2^{pp})\over \epsilon_2^{pp}})^2 
 \nonumber \\
&&=0.5 : 0.3: 0.16 
\ea
(This is not inconsistent with constancy of the $v_2/\epsilon_2$ proposed above, since in the calculations we do not compare three points on
the same adiabatic or $RT=const$.)

The CMS data do show that the pp has smaller $v_2$ as compared to $pA$ data, the ratio is about a
factor of 1/4 (see Fig.3 of  \cite{CMS:2012qk}) rather than
1/2 which the hydro solution provides. Perhaps it is because the pp collisions create a somewhat more spherical
fireball, with
$ \epsilon_2^{pp} <  \epsilon_2^{pA} $,  
 in spite of having a smaller size. We will return to this issue at the end of the paper.

Let us now compare in a similar manner the ratio of the $m=3$ to $m=2$ harmonics 
\ba 
&&\left({v_3^{AA} \over v_2^{AA}  }\right)^2\approx 0.12\left({\epsilon_3^{AA} \over \epsilon_2^{AA}  }\right)^2  \\ \nonumber
&&\left({v_3^{pA} \over v_2^{pA}  }\right)^2\approx 0.09 \left({\epsilon_3^{pA} \over \epsilon_2^{pA}  }\right)^2  \\ \nonumber
&&\left({v_3^{pp} \over v_2^{pp}  }\right)^2\approx 0.02 \left({\epsilon_3^{pp} \over \epsilon_2^{pp}  }\right)^2 
\ea
Assuming $\epsilon_3/\epsilon_2\sim 1$ one finds that  in pA we predict  
$v_3/v_2\approx 1/3$, which agrees nicely with the ALICE data  \cite{Abelev:2012cya}.
For pp  we have $v_3/v_2\approx 1/7$
which is probably too small to be seen.


\subsection{Comment of higher gradients at freezeout} 

The effect of flow gradients affect spectra at freezeout.
As emphasized by
Teaney \cite{Teaney:2003kp}, the  equilibrium  distribution function $f_0(x,p)$ 
should be complemented  by the non equilibrium corrections proportional to flow gradients 
\ba f(x, p) =&&f_0(x, p) + \delta f(x,p)  p^\mu p^\nu \partial_\mu u_\nu\nonumber\\
&& + ({\rm higher \,\,gradients}) 
\ea
Furthermore \cite{Teaney:2003kp} , the Lorentz covariance
forces any extra gradient to carry another power of the particle momentum. As a result, 
the expansion parameter of the n-th term is of the order
\be {\delta f \over f}\sim  {\eta\over s} \left( {p\over T} {1 \over T R} \right)^n\ee
If one moves to large $p_\perp/T= {\cal O}(10)$, compensating small factor $1/TR$, the expansion in gradients 
(and thus hydrodynamics)  breaks down. Indeed, the radial and  harmonics of the flow
agree with hydro up to  transverse  momenta of the order of  $p_t\approx 3 \, {\rm GeV}$, or $p_t/T_f < 20$.

In some applications people had calculated $f+\delta f$ and get negative spectra at large $p_t$
from viscous corrections, which is of course meaningless.  Needless to say, it resemble the first terms $(1-x)$
 in expansion (\ref{expansion}), which gets negative for $x>1$. Our suggestion, along the line of LS re-summation,
 is to use instead
 \be f= {f_0 \over 1+ \delta f/f_0} \ee
form which is sign-definite and approximately reproduce the higher order terms as well. 

%
%



 \section{Summary and Discussion} 
High multiplicity pp and pA collisions are very interesting systems to study, as they are expected to display the transition
from a ``micro" to ``macro" dynamical regimes, treated theoretically by quite different means. In this paper we tried to explain
how this transition works using the language of the macroscopic theory, the viscous hydrodynamics.

As we emphasized in the Introduction, the applicability of hydrodynamics to high energy collisions rests on {\em the product of the two} 
small parameters:  (i)  the  micro-to-macro ratio $1/TR$,  and (ii) the viscosity-to-entropy ratio $\eta/s$. For central AA collisions,
both are small or of order ${\cal O}(1/10) $. For  high enough multiplicity of the pA and pp collisions, such as the first  parameter 
becomes the same as in current AA collisions, the accuracy of hydrodynamics should be the same.
While those value of multiplicity are not reached yet, hydrodynamics apparently starts to work, albight
 with less accuracy.

After solving the hydrodynamical equations we found that the {\em radial} (axially symmetric) flow is 
little modified by viscosity and is in fact enhanced by ``longer" (in dimensionless time) run.
Thus our main prediction is an {\em enhanced radial flow}. Its  signatures 
--  growing $m_\perp$ slopes with the particle mass, or growing
    proton-to-pion-ratio -- are indeed confirmed by recent CMS and ALICE
    data. This happens in spite of the fact, that AA feezeout happens at smaller $T_f$ than in pA.
    
    We extensively studied various forms of viscous hydrodynamics, from NS to IS to re-summation of gradients as la LS.
  In short, those grow  from AA to pA to pp, but perhaps  even in the last case they remain manageable.  
    Higher harmonics are obviously more penalized by viscous corrections, especially of higher order, as each
    gradient goes with extra factor $m$. The role of those should be studied further elsewhere.
 
    Finally let us comment the following: version 1 of this paper also included a view on the high multiplicity pp/pA from
 microscopic model, based on stringy Pomeron. It had grown substantially and will now appear as a separate publication.
    
 {\bf Note added:} When this version of the paper was completed, we learned about ALICE measurements
 of the identified particle spectra in high multiplicity pPb collisions \cite{Abelev:2013haa}.  
 Strong radial flow, growing with the multiplicity, is reported, clearly seen in proton/antiproton spectra.
All conclusions are completely consistent with ours. Note especially one point: ALICE also finds that in
pPb the freezeout  happens at temperature $T_f^{pPb}\approx 0.17\, GeV$
higher than that in central PbPb, in which  $T_f^{PbPb}\approx 0.12\, GeV$. 
  \vskip 1cm
{\bf Acknowledgements.} ES would like to thank  former student Pilar Staig for her help in developing many of the ideas
presently discussed. We also thank Gokce Basar, Dima Kharzeev and Derek Teaney for discussions.
This work was supported by the U.S. Department of Energy under Contract No. DE-FG-88ER40388.

\end{document}